\documentclass[10pt,british,two columns]{IEEEtran}
\usepackage[T1]{fontenc}
\usepackage[latin9]{inputenc}
\usepackage{mathtools}
\usepackage{bm}
\usepackage{amsmath}
\usepackage{amsthm}
\usepackage{amssymb}
\usepackage{graphicx}

\makeatletter
\theoremstyle{plain}
\newtheorem{thm}{\protect\theoremname}
\theoremstyle{plain}
\newtheorem{lem}[thm]{\protect\lemmaname}
\theoremstyle{plain}
\newtheorem{prop}[thm]{\protect\propositionname}

\usepackage{cite}

\@ifundefined{showcaptionsetup}{}{%
 \PassOptionsToPackage{caption=false}{subfig}}
\usepackage{subfig}
\makeatother

\usepackage{babel}
\providecommand{\lemmaname}{Lemma}
\providecommand{\propositionname}{Proposition}
\providecommand{\theoremname}{Theorem}

\begin{document}
\title{Optimal Number of Measurements in a Linear System with Quadratically
Decreasing SNR \thanks{This paper was presented in part at 25th European Signal Processing
Conference \cite{lu2017optimal}. }\thanks{This work is partially supported by US ONRG Grant N62909-16-1-2052.}}
\author{Yang Lu\textsuperscript{1}, Wei Dai\textsuperscript{1} and Yonina
C. Eldar\textsuperscript{2}\\
{\small{}}\textsuperscript{1}{\small{}Department of Electrical and
Electronic Engineering, Imperial College London, United Kingdom}\linebreak{}
{\small{}}\textsuperscript{2}{\small{}Department of Mathematics
and Computer Science, Weizmann Institute of Science, Rehovot, Israel.}}
\maketitle
\begin{abstract}
We consider the design of a linear sensing system with a fixed energy
budget assuming that the sampling noise is the dominant noise source.
The energy constraint implies that the signal energy per measurement
decreases linearly with the number of measurements. When the maximum
sampling rate of the sampling circuit is chosen to match the designed
sampling rate, the assumption on the noise implies that its variance
increases approximately linearly with the sampling rate (number of
measurements). Therefore, the overall SNR per measurement decreases
quadratically in the number of measurements. Our analysis shows that,
in this setting there is an optimal number of measurements. This is
in contrast to the standard case, where noise variance remains unchanged
with sampling rate, in which case more measurements imply better
performance. Our results are based on a state evolution technique
of the well-known approximate message passing algorithm. We consider
both the sparse (e.g. Bernoulli-Gaussian and least-favorable distributions)
and the non-sparse (e.g. Gaussian distribution) settings in both the
real and complex domains. Numerical results corroborate our analysis.
\end{abstract}

\begin{IEEEkeywords}
Approximate message passing, compressed sensing, state evolution,
signal recovery.
\end{IEEEkeywords}

\section{Introduction}

The problem of estimating a signal from its linear measurements has
been studied for many decades. From linear algebra, at least $n$
measurements are required in order to ensure the reconstruction of
an $n$-dimensional signal. Otherwise, the solution is not unique.
The field of compressed sensing (CS) \cite{Eldar1,Eldar2} has shown
that when the unknown signal is sparse, assuming no further prior
knowledge on the signal, the number of measurements can be reduced
below the dimension of the signal leading to an underdetermined linear
measurement system. Many low complexity algorithms have been proposed
to solve the resulting sparse recovery problem including greedy algorithms,
for example orthogonal matching pursuit (OMP) \cite{342465}, subspace
pursuit (SP) \cite{4839056} and compressive sampling matching pursuit
(CoSaMP) \cite{needell2009cosamp}, $\ell_{1}$-norm minimisation
\cite{tibshirani1996regression}, and more recently approximate message
passing (AMP) \cite{donoho2009message}. CS has been widely used in
under-sampling \cite{8004458EldarSub,7883952EldarSub,donoho2009message,OnJointRecoveryXiaoChen},
imaging and localisation \cite{baraniuk2007compressiveRadar,5393298OFDMradar,AMPimaging},
and sparse learning \cite{AMP_sparse_learning}. 

In this paper, we study a system design problem with focus on the
sampling rate. That is, based on the typical characteristics of the
acquired signals, our goal is to choose an optimal number of measurements
to minimise the mean squared error (MSE) of the recovery. We assume
that the signal's dimension, sparsity level, and statistics are given.
We also make two assumptions on the sampling process. First, since
practical systems are power limited, the energy in the measurements
for a fixed time period is fixed. This implies that the energy per
measurement decreases linearly with the number of measurements (or
equivalently sampling rate). Second, we assume that the sampling noise
is the dominant noise source. In addition, in order to minimise hardware
cost, the sampling circuit is chosen to operate at its maximum sampling
rate. We model the sampling noise as additive white Gaussian noise.
The noise variance of the sampling circuit follows the well known
$KT/C$ rule \cite{4052064,5669444design,razavi2001design}, where
$K$ is the Boltzmann's constant, $T$ is the absolute temperature
of the circuit, and $C$ is the capacitance of the sampling circuit
which is inversely proportional to the maximum and optimal sampling
rate based on our assumption. This implies that the noise variance
increases approximately linearly with the number of measurements.
In addition, although quantisation noise is not studied in this paper,
it is widely observed that the effective number of bits of an analog-to-digital
converter decreases when the sampling frequency increases \cite{davenport2012pros}.
This also results in the phenomenon that higher sampling rate implies
larger noise variance. The combined effect of these two assumptions
is that the SNR per measurement decreases quadratically as a function
of the number of measurements.

In practical systems, other noise sources exist such as additive noise
before sampling (also known as signal noise) \cite{arias2011noise,davenport2012pros}
and quantisation noise\cite{laska2012regime,5773638}. In this paper,
we focus on the effect of sampling noise, leaving analysis of the
effects of other noises as possible future work.

Under a quadratically decreasing SNR system, our goal is to analyse
the optimal number of measurements. Different from the standard setting
where noise variance remains unchanged with sampling rate and hence
more measurements typically means better recovery performance, we
show that with quadratically decreasing SNR more measurements do not
necessarily imply better recovery.

More specifically, we demonstrate that in the quadratically decreasing
SNR scenario, there exists an optimal normalised number $\delta^{\dagger}$
of measurements to minimise the mean squared error (MSE) of the recovered
signal. Here $\delta=\frac{m}{n}$, where $m$ is the number of measurements
and $n$ is the dimension of the unknown signal. We explicitly study
three signal models: Gaussian, Bernoulli-Gaussian and least-favorable
distributions in both the real and complex domains. We obtain closed-form
expressions for $\delta^{\dagger}$ in the Gaussian and least-favorable
models and provide a numerical procedure to find the optimal $\delta^{\dagger}$
in other cases. We show that for the three models, $\delta^{\dagger}\leq2$.
Furthermore, when the SNR is low, $\delta^{\dagger}$ can be smaller
than $1$ implying that $m<n$ in both the sparse and non-sparse models.
In particular, $\delta^{\dagger}<1$ in the Gaussian model when $\sigma_{x}^{2}<2\sigma_{0}^{2}$
where $\sigma_{x}^{2}$ is the signal variance and $\sigma_{0}^{2}$
is the noise base level, defined in Section \ref{subsec:System-Model}.
For sparse vectors we require in addition that $\epsilon=\frac{S}{n}$
is smaller than roughly $0.2$ where $S$ is the number of non-zero
elements in the unknown signal.

Our analysis and results are based on the AMP algorithm and the associated
state evolution. Though rigorous derivation of the state evolution
of AMP requires a random Gaussian matrix, many works have demonstrated
that the same results are relatively accurate for partial Fourier
and Rademacher matrices \cite{donoho2009message,donoho2009supporting}
when the sizes of these matrices are sufficiently large.

The rest of this paper is organised as follows: In Section \ref{sec:Preliminaries}
we introduce our problem and mathematically explain the quadratically
decreasing SNR model. We also provide the relationship between $\delta^{\dagger}$
and the MSE in the Gaussian setting based on random matrix theory.
In Section \ref{sec:Background-of-Approximate}, we introduce AMP
and state evolution which are used for sparse recovery. The analysis
of least-favorable and Bernoulli-Gaussian models is developed in Section
\ref{sec:Analysis-in-Real} for the real case and in Section \ref{sec:Analysis-in-Complex}
for the complex case. Bounds on the optimal number of measurements
and discussion about the specific situation in which $\delta^{\dagger}<1$
are provided in Section \ref{sec:BoundsANDSimulation}, followed by
conclusions.

\section{\label{sec:Preliminaries}System Model}

\subsection{\label{subsec:Quadratic-Decreasing-SNR}Quadratically Decreasing
SNR Model}

Consider the linear system
\begin{equation}
\bm{y}=\bm{s}+\bm{w},\label{eq:linear system}
\end{equation}
where $\bm{y}\in\mathbb{H}^{m}$ is the observation vector, $\bm{s}\in\mathbb{H}^{m}$
represents an unknown signal vector and $\bm{w}\in\mathbb{H}^{m}$
is additive white Gaussian noise. Here, $\mathbb{H}\in\left\{ \mathbb{R},\mathbb{C}\right\} $
where $\mathbb{R}$ denotes the real domain and $\mathbb{C}$ the
complex domain.

Assume the system has a fixed energy budget $E_{s}$, so that the
total energy that can be allocated to $\bm{s}$ is
\begin{equation}
E_{s}=\sum_{i=1}^{m}\left|s_{i}\right|^{2}.\label{eq:fixed total energy}
\end{equation}
The corresponding average energy per signal sample is
\begin{align}
P_{as} & =\frac{1}{m}\sum_{i=1}^{m}\left|s_{i}\right|^{2}=\frac{1}{m}E_{s}\propto\frac{1}{m}.\label{eq:average energy}
\end{align}

\begin{figure}
\centering{}\includegraphics[scale=0.48]{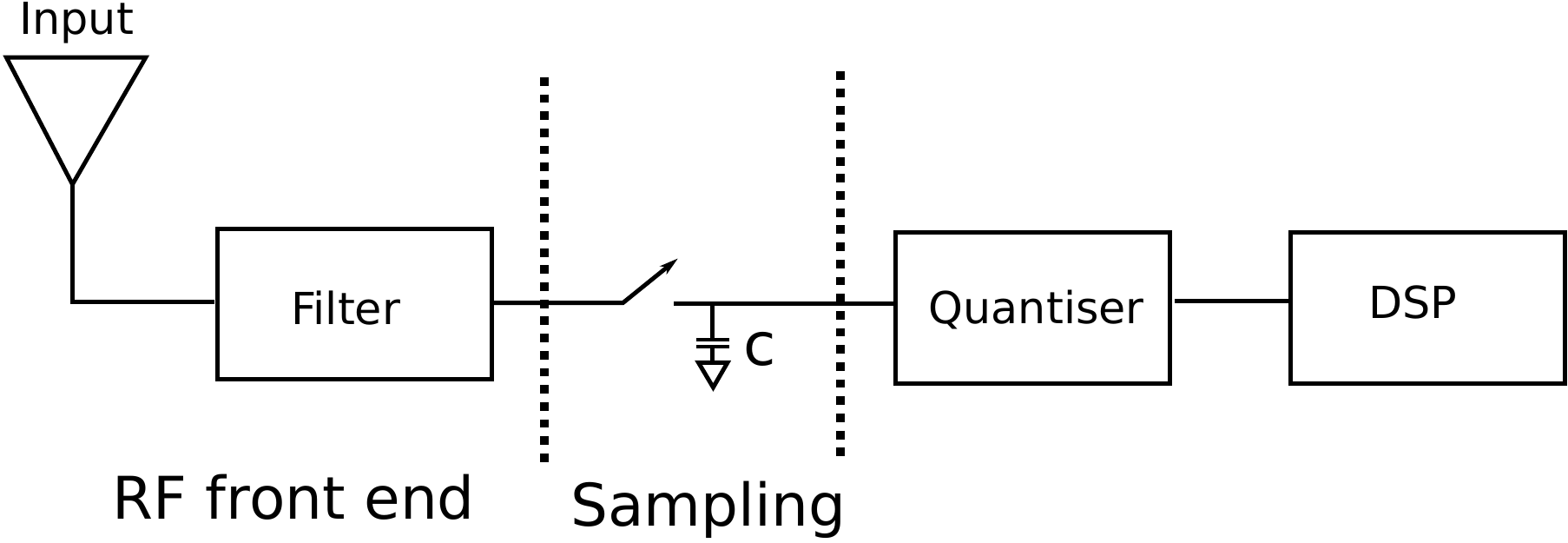}\caption{\label{fig:Signal-acquisition-system}Signal acquisition system.}
\end{figure}

In practice, noise is unavoidable and comes from different parts of
the system. Consider the signal acquisition scheme illustrated in
Fig. \ref{fig:Signal-acquisition-system}. The received analog signal
first passes through a low pass or band pass filter, which filters
the out of band noise. We refer to the remaining in-band noise as
signal noise. The filtered signal is then sampled which produces sampling
noise. After that, a quantiser is applied to convert the samples to
bits which leads to quantisation errors. In \cite{arias2011noise,davenport2012pros,laska2012regime},
the authors discuss the signal noise and the corresponding noise folding
effect. Quantisation error was studied in \cite{5773638}. A combined
analysis of the signal noise and quantisation errors is provided in
\cite{laska2012regime} with the assumption of a fixed bit-budget
on the measurements. 

The sampling noise (i.e. thermal noise in the sampling circuit) is
often the dominant noise source, for example in radar systems \cite[pp 5.21-5.26]{NAWCWDTP8347}
and magnetic resonance imaging (MRI) systems \cite{aja2013review}.
Here we focus on the impact of sampling noise by assuming that both
signal noise and quantisation errors are relatively small. Based on
\cite{4052064,5669444design,razavi2001design}, the variance of sampling
noise is approximately given by $KT/C$ where $K$ is the Boltzmann's
constant, $T$ is the absolute temperature of the circuit and $C$
is the sampling capacitance. The speed of the sampling circuit (which
can be simplified as a switched-capacitor circuit) depends on the
on-resistance of the switch and the sampling capacitance $C$. A high
speed sampling circuit requires a small sampling capacitor. From the
system design point of view, there is a trade-off between the sampling
noise (with variance $KT/C$) and the designed sampling speed. This
phenomenon is discussed in Chapter 12 of \cite{razavi2001design}
in detail. Here we emphasise that the sampling noise in the sampling
circuit is independent of the bandwidth of the signal $\bm{s}$ and
only depends on the sampling rate of the circuit.

As discussed above, the noise variance increases approximately linearly
with the number of measurements. The average energy per noise sample
is then
\begin{align*}
P_{an} & \propto f_{s}\propto m
\end{align*}
where $f_{s}$ is the sampling frequency. With these definitions,
the SNR is proportional to
\[
{\rm SNR}=\frac{P_{as}}{P_{an}}\propto\frac{\frac{1}{m}}{m}=\frac{1}{m^{2}}
\]
which is a quadratic decreasing model with respect to the number of
measurements $m$. 

We have implicitly assumed that the signal noise and the quantisation
error are negligible compared to the sampling noise \cite{NAWCWDTP8347,aja2013review}.
In reality, the former two noises will create a small but nonzero
noise floor with constant variance. As the variance of the noise floor
depends on system specifications, we leave the detailed design considering
all noise sources as a future research topic.

\subsection{\label{subsec:System-Model}General Linear System Model}

We now extend the model in (\ref{eq:linear system}) to a general
linear model
\begin{equation}
\bm{y}=\bm{A}\bm{x}+\bm{w}\label{eq:linear system 1}
\end{equation}
where $\bm{A}\bm{x}\coloneqq\bm{s}$, $\bm{A}\in\mathbb{H}^{m\times n}$
denotes the measurement matrix and $\bm{x}$ represents the unknown
signal vector. In order to incorporate the quadratically decreasing
SNR assumption, we define $\bm{A}$ as a Gaussian random matrix with
elements i.i.d. drawn from $\mathcal{N}\left(0,\frac{1}{m}\right)$
or $\mathcal{CN}\left(0,\frac{1}{m}\right)$. The elements of $\bm{x}$
are assumed to be i.i.d. drawn from a distribution $p_{x}$ with zero
mean and variance independent of $m$. We assume the noise is Gaussian
$\mathcal{N}\left(0,\sigma_{w}^{2}\right)$ or $\mathcal{CN}\left(0,\sigma_{w}^{2}\right)$
with
\begin{equation}
\sigma_{w}^{2}\coloneqq\delta\sigma_{0}^{2}\label{eq:noise model}
\end{equation}
where $\delta\coloneqq\frac{m}{n}$ and $\sigma_{0}^{2}$ is the constant
noise base level.

Let $\hat{\bm{x}}$ be an estimate of $\bm{x}$. The associated MSE
is defined by
\begin{align}
{\rm Err} & =\lim_{n\rightarrow\infty}\frac{1}{n}{\rm {\rm \mathbb{E}}}\left[\left\Vert \bm{x}-\hat{\bm{x}}\right\Vert ^{2}\right].\label{eq:MSE}
\end{align}
Under the quadratically decreasing SNR model we want to determine
the optimal number of measurements $m$ to minimise the MSE in estimating
$\bm{x}$. In particular, an interesting question is whether there
are cases where $m$ is smaller than $n$, which would imply an under-sampling
scenario. We consider the system model (\ref{eq:linear system 1})
for both non-sparse and sparse signals.

\subsection{\label{subsec:Non-Sparse-Setting}Non-Sparse Setting}

To gain intuition, we start by considering a Gaussian model. Assume
that $\bm{x}$ is drawn from {\small{}$\mathcal{N}\left(\bm{0},\sigma_{x}^{2}\bm{I}\right)$
when} $\mathbb{H}=\mathbb{R}$ (or{\small{} $\mathcal{CN}\left(\bm{0},\sigma_{x}^{2}\bm{I}\right)$
when }$\mathbb{H}=\mathbb{C}$) and the noise $\bm{w}$ is drawn from
{\small{}$\mathcal{N}\left(\bm{0},\sigma_{w}^{2}\bm{I}\right)$ when}
$\mathbb{H}=\mathbb{R}$ (or{\small{} $\mathcal{CN}\left(\bm{0},\sigma_{w}^{2}\bm{I}\right)$
when }$\mathbb{H}=\mathbb{C}$). The asymptotic MSE (\ref{eq:MSE})
of the MMSE estimator can then be directly calculated based on random
matrix theory \cite{5695122}. Denoting $c=\frac{\left(1-\delta\right)}{\delta}$,
we have
\begin{equation}
{\rm Err}=\!\frac{\delta}{2}\left[\left(\!-\!\sigma_{w}^{2}\!\!+\!c\sigma_{x}^{2}\right)\!\!+\!\!\sqrt{\!\left(\sigma_{w}^{2}\!\!+\!c\sigma_{x}^{2}\right)^{2}\!\!+\!4\sigma_{w}^{2}\sigma_{x}^{2}}\right].\label{eq:RandomMatrixTheory_Gaussian}
\end{equation}

\begin{figure}
\begin{centering}
\subfloat[\label{fig:Traditional-Case}Traditional Case: $\sigma_{w}^{2}=\sigma_{0}^{2}$
is constant]{\begin{centering}
\includegraphics[scale=0.22]{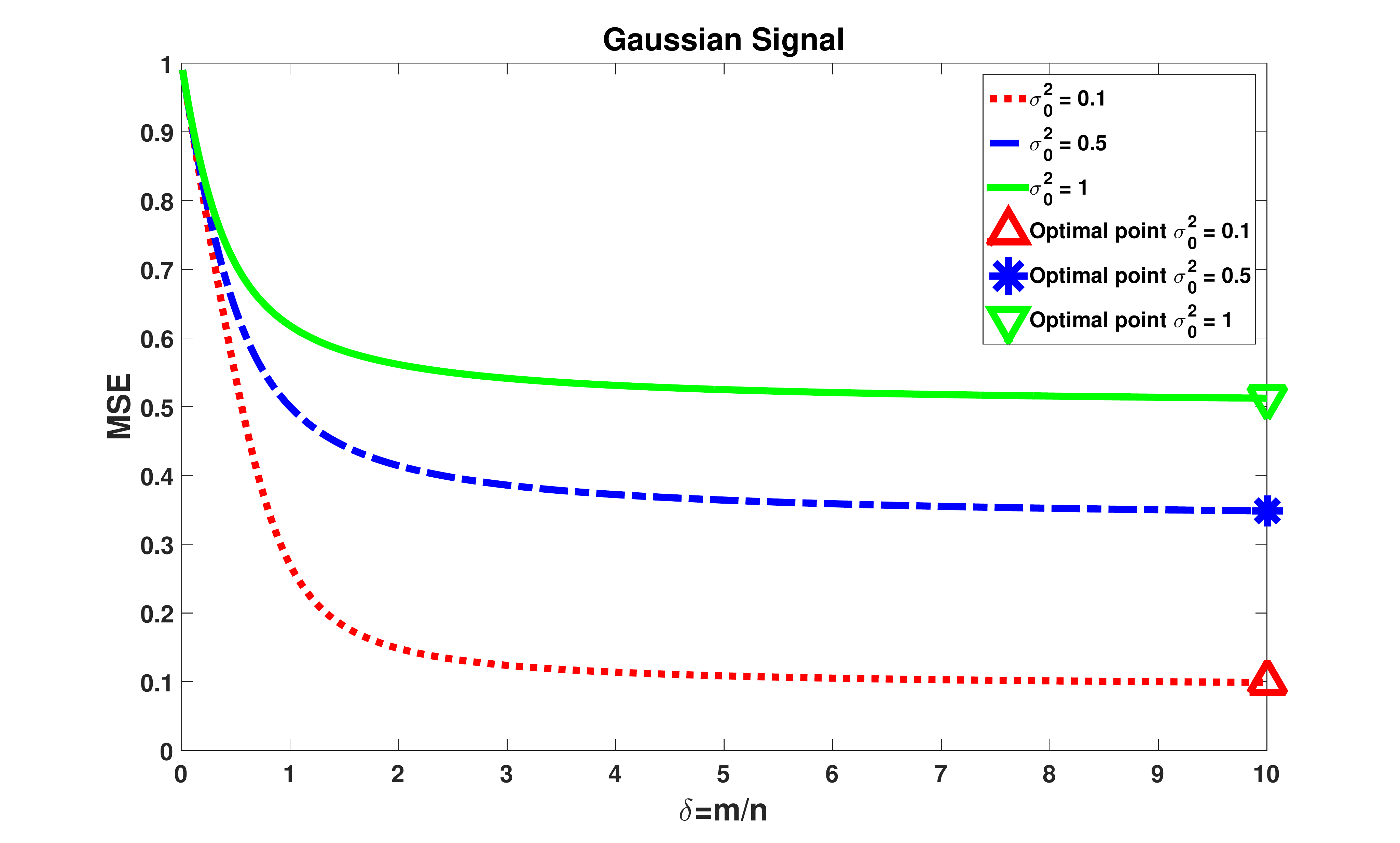}
\par\end{centering}
}
\par\end{centering}
\centering{}\subfloat[\label{fig:Our-Case}Our Case: $\sigma_{w}^{2}=\delta\sigma_{0}^{2}$
varies with $\delta$]{\begin{centering}
\includegraphics[scale=0.22]{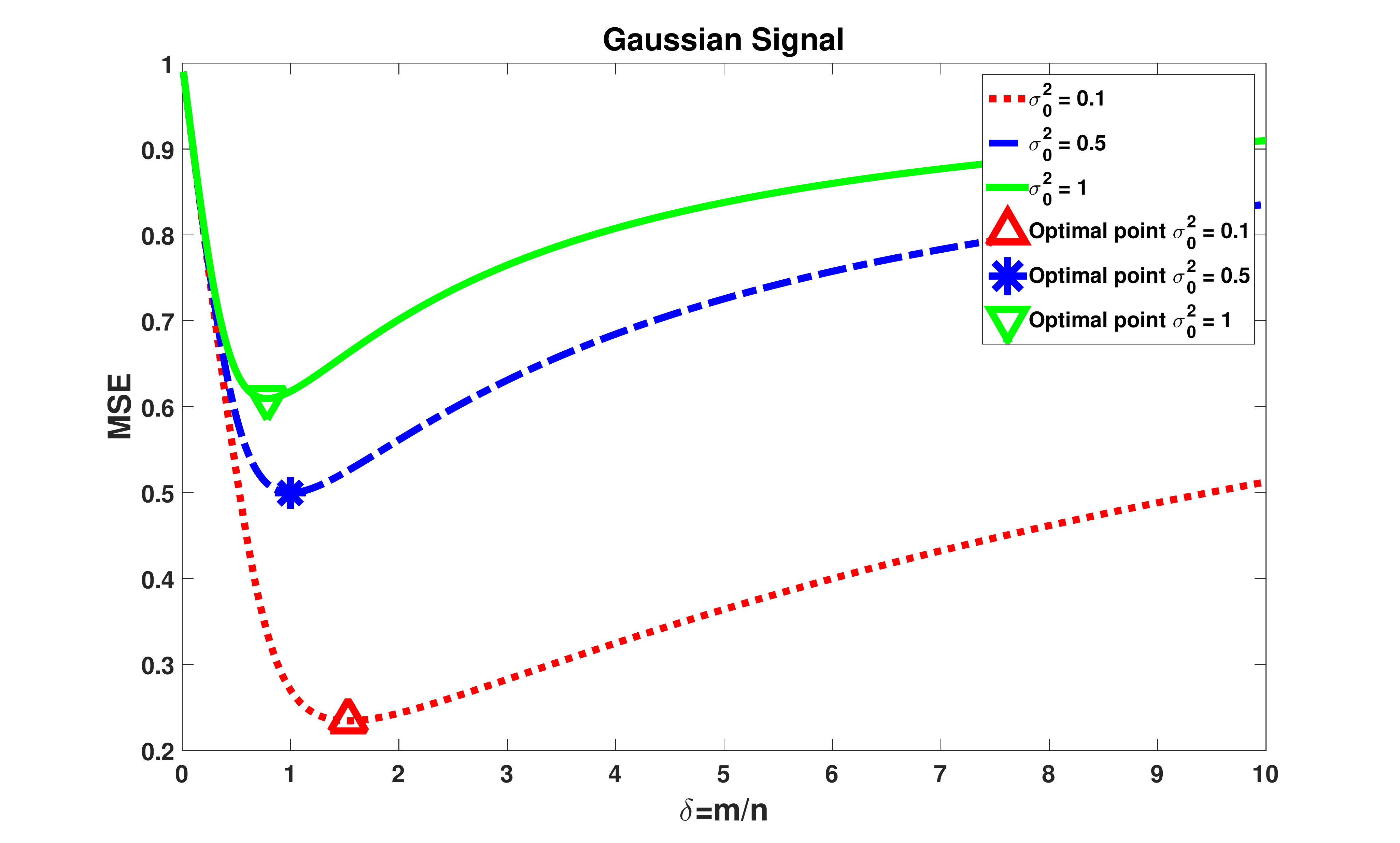}
\par\end{centering}
}\caption{\label{fig:Gaussian-Signals}MSE in a non-sparse Gaussian model.}
\end{figure}
An illustration of the MSE for the case in which $\sigma_{x}^{2}=1$
and $\sigma_{w}^{2}=\sigma_{0}^{2}$ is provided in Fig. \ref{fig:Traditional-Case}.
The MSE is calculated via (\ref{eq:RandomMatrixTheory_Gaussian})
and we choose three different values for $\sigma_{0}^{2}$: $0.1$,
$0.5$ and $1$. The simulation results reflect, in the traditional
case, that more measurements provide better performance. When we replace
the noise variance with our model (\ref{eq:noise model}), a trade-off
between MSE and $\delta$ exists. Figure \ref{fig:Our-Case} shows
that there is an optimal number of measurements that minimizes the
MSE which is quit different from the traditional case. The optimal
values of $\delta$ for the three curves ($\sigma_{0}^{2}=0.1$, $0.5$
and $1$) are smaller than $2$; when $\sigma_{0}^{2}=1$, the optimal
$\delta$ is even smaller than $1$ (i.e. $m<n$). A detailed discussion
regarding bounds on $\delta^{\dagger}$ and the situation in which
$\delta^{\dagger}<1$ is provided in Section \ref{sec:Boundary-Analysis}.

Motivated by this phenomenon, our goal is to find a similar relationship
for sparse signals by taking into account the nonlinear property of
the sparse decoder.

\section{\label{sec:Background-of-Approximate}Approximate Message Passing}

In order to treat sparse signals, a sparse decoder is required for
signal recovery. We choose the AMP algorithm which has low computational
complexity per iteration, fast convergence speed (when it converges),
and good performance guarantees (with a standard Gaussian random matrix).
Exact performance of AMP can be predicted via the so called state
evolution (SE) technique.

The AMP algorithm was first proposed in \cite{donoho2009message}
to solve (\ref{eq:linear system 1}) in the CS scenario in which $m<n$
and $\bm{x}$ is assumed to be sparse. It iteratively applies the
following equations:
\begin{eqnarray}
\bm{x}^{t+1} & =\! & \eta\left(\bm{A}^{*}\bm{r}^{t}+\bm{x}^{t}\right),\label{eq:AMP1}\\
\bm{r}^{t} & =\! & \bm{y}\!-\!\bm{A}\bm{x}^{t}+\frac{1}{\delta}\!\left\langle \eta^{\prime}\!\left(\bm{A}^{*}\bm{r}^{t-1}\!+\!\bm{x}^{t-1}\right)\!\right\rangle \!\bm{r}^{t-1},\label{eq:AMP2}
\end{eqnarray}
where $\bm{x}^{t}$ denotes the $t$-th estimation of $\bm{x}$, $\eta\left(\cdot\right)$
is a component-wise estimator designed based on the statistical information
on its input argument, $\bm{A}^{*}$ stands for the (conjugate) transpose
of $\bm{A}$, $\eta^{\prime}$ represents the first order derivative
of $\eta$, and $\left\langle \bm{v}\right\rangle :=\frac{1}{n}\sum_{i=1}^{n}v_{i}$
computes the average. The last term of (\ref{eq:AMP2})
\begin{equation}
{\rm Onsager}\coloneqq\frac{1}{\delta}\left\langle \eta^{\prime}\left(\bm{A}^{*}\bm{r}^{t-1}+\bm{x}^{t-1}\right)\right\rangle \bm{r}^{t-1},\label{eq:Onsager Term}
\end{equation}
is called the Onsager (correction) term.

A heuristic suggestion was presented in \cite{5695122,montanari2012graphical}
to analyse the performance of AMP. The idea is to decompose the input
of the function $\eta\left(\bm{\beta}^{t}\right)$ into the superposition
of the original signal $\bm{x}$ and white Gaussian noise. Consider
three modifications at each iteration $t$: replace 1) the matrix
$\bm{A}$ with a new independent copy $\bm{A}\left(t\right)$, 2)
the observation vector $\bm{y}$ with $\bm{y}^{t}=\bm{A}\left(t\right)^{*}\bm{x}+\bm{w}$,
and 3) the Onsager term in (\ref{eq:AMP2}) with $\bm{0}$. The input
of $\eta\left(\bm{\beta}^{t}\right)$ can then be written as
\begin{align}
\bm{\beta}^{t} & \coloneqq\bm{A}\left(t\right)^{*}\bm{r}^{t}+\bm{x}^{t}=\bm{A}\left(t\right)^{*}\left(\bm{y}^{t}-\bm{A}\left(t\right)\bm{x}^{t}\right)+\bm{x}^{t}\nonumber \\
 & =\bm{x}+\left(\bm{A}\left(t\right)^{*}\bm{A}\left(t\right)-\bm{I}\right)\left(\bm{x}-\bm{x}^{t}\right)+\bm{A}\left(t\right)^{*}\bm{w}\label{eq:equivalent signal}
\end{align}
which is the ground truth signal $\bm{x}$ plus an equivalent noise:
\begin{equation}
\bm{w}_{e}^{t}\coloneqq\left(\bm{A}\left(t\right)^{*}\bm{A}\left(t\right)-\bm{I}\right)\left(\bm{x}-\bm{x}^{t}\right)+\bm{A}\left(t\right)^{*}\bm{w}.\label{eq:equivalent noise}
\end{equation}

Although in the AMP algorithm, we do not have independent copies $\bm{A}\left(t\right)$
as well as independent observations $\bm{y}^{t}$, due to the Onsager
term, the statistical information of the equivalent noise can still
be asymptomatically calculated via (\ref{eq:equivalent noise}) which
may be treated as approximately Gaussian and independent of $\bm{x}$.
At each iteration $t$, we first update the estimated signal $\bm{x}^{t+1}=\eta\left(\bm{\beta}^{t}\right)$
by a particularly designed $\eta\left(\cdot\right)$ function (detailed
information is provided in Sections \ref{sec:Analysis-in-Real} and
\ref{sec:Analysis-in-Complex}) which depends on the knowledge of
the signal $\bm{x}$ and the equivalent noise $\bm{w}_{e}^{t}$ (i.e.
$\sigma_{e}^{t}$, the standard deviation of $\bm{w}_{e}^{t}$). We
then calculate the MSE denoted by ${\rm Err}_{t+1}$ of the current
estimated signal $\bm{x}^{t+1}$. Finally we update the variance of
the equivalent noise $\bm{w}_{e}^{t+1}$ of (\ref{eq:equivalent noise})
by
\begin{equation}
\left(\sigma_{e}^{t+1}\right)^{2}=\frac{1}{\delta}{\rm Err}_{t+1}+\sigma_{w}^{2}.\label{eq:SE_COMMON}
\end{equation}
The calculation of ${\rm Err}_{t+1}$ depends on the distribution
of $\bm{x}$ and will be discussed in Sections \ref{sec:Analysis-in-Real}
and \ref{sec:Analysis-in-Complex}. This updated variance of $\bm{w}_{e}^{t+1}$
is used to obtain a new signal estimate $\bm{x}^{t+2}$ in the next
iteration.

The asymptotic performance of the AMP algorithm in the regime in which
$m,\,n\rightarrow\infty$ with $\delta\rightarrow\frac{m}{n}$ constant,
is described by SE. The SE is characterised by a sequence $\{(\sigma_{e}^{t})^{2}\}$
for $t\geq0$ calculated via (\ref{eq:SE_COMMON}) with initial condition
$\left(\sigma_{e}^{0}\right)^{2}={\rm \mathbb{E}}\left[X^{2}\right]/\delta+\sigma_{w}^{2}$
($X$ has density function $p_{x}$). As long as $\left(\sigma_{e}^{t+1}\right)^{2}\leq\left(\sigma_{e}^{t}\right)^{2}$
for all $t\geq0$, we say that AMP converges. More details about SE
can be found in \cite{montanari2012graphical,5695122}. The analysis
of the optimal number of measurements to minimise the MSE in this
paper assumes that AMP converges and relies heavily on the SE.

In the rest of this paper, when we say the practical performance of
AMP, we refer to the practical situation in which $n$ is a large
but finite number. We iteratively apply (\ref{eq:AMP1}) and (\ref{eq:AMP2})
to update the estimate $\bm{x}^{t}$. In each iteration, the MSE (6)
can be approximated by $\frac{1}{n}\left\Vert \bm{x}-\bm{x}^{t}\right\Vert ^{2}$.
We are interested in finding $\frac{1}{n}\left\Vert \bm{x}-\bm{x}^{\infty}\right\Vert ^{2}$,
where $\bm{x}^{\infty}$ represents the final estimate of $\bm{x}$
when AMP converges. When we say the theoretical performance of AMP,
we refer to SE analysis via (\ref{eq:SE_COMMON}). In this case, the
corresponding MSE is ${\rm Err}_{\infty}$ when AMP converges. An
advantage of AMP is that when it converges, the theoretical value
${\rm Err}_{\infty}$ describes the practical performance calculated
by $\frac{1}{n}\left\Vert \bm{x}-\bm{x}^{\infty}\right\Vert ^{2}$
quite precisely. The performance guarantee of AMP has been rigorously
analysed both in the infinite \cite{5695122} and finite \cite{8318695}
domain.

\section{\label{sec:Analysis-in-Real}Analysis in Real Domain}

We now analyse the relationship between the MSE and $\delta$ (or
equivalently $m$) in the real domain. We first consider a situation
in which the only prior information on the unknown signal is that
it is sparse. Because the detailed distribution is not available,
a universal decoder is applied and analysed based on the so-called
least-favorable distribution. Then we treat the case in which the
signal distribution is known a priori and given by Bernoulli-Gaussian.
We extend the analysis to the complex domain in the next section.

\subsection{\label{subsec:Least-favorite-Distribution}Least-favorable Distribution
(Worst Case Analysis)}

Consider a vector $\bm{x}$ with i.i.d. elements drawn from an unknown
distribution $p_{x}$ supported on $\left(-\infty,\infty\right)$,
where only the normalised sparsity level $\epsilon=\frac{S}{n}$ is
given. Denote the class of corresponding signals as $\mathcal{F}_{\epsilon}$,
such that $p_{x}\in\mathcal{F}_{\epsilon}$. In \cite{donoho2009message,montanari2012graphical}
a soft-thresholding function is used component-wise as the estimato{\small{}r}
\begin{equation}
x_{i}^{t+1}=\eta\left(\beta_{i}^{t},\lambda^{t}\right)=\begin{cases}
\beta_{i}^{t}-\lambda^{t} & {\rm }\,\beta_{i}^{t}>\lambda^{t}\\
0 & {\rm }\,-\lambda^{t}\leq\beta_{i}^{t}\leq\lambda^{t}\\
\beta_{i}^{t}+\lambda^{t} & {\rm }\,\beta_{i}^{t}<-\lambda^{t}.
\end{cases},\label{eq:soft-threshold-function}
\end{equation}
In AMP, $\bm{\beta}^{t}$ is calculated by $\bm{A}\left(t\right)^{*}\bm{r}^{t}+\bm{x}^{t}$
with initial condition $\bm{r}^{0}=\bm{y}$ and $\bm{x}^{0}=\bm{0}$,
and the non-negative value $\lambda^{t}$ is the corresponding threshold
which depends on the equivalent noise $\bm{w}_{e}^{t}$ of (\ref{eq:equivalent noise}).
The selection of $\lambda^{t}$ is detailed in (\ref{eq:optimal_alpha})
below.

The AMP state evolution is based on the following component-wise analysis.
Let $\hat{X}$ be an estimate of a random variable $X$. The worst
case analysis considers the following minimax problem
\[
\inf_{\hat{X}}\sup_{p_{x}\in\mathcal{F}_{\epsilon}}{\rm \mathbb{E}}\left[\left|X-\hat{X}\right|^{2}\right],
\]
which minimises the MSE under the least-favorable distribution. When
the estimator (\ref{eq:soft-threshold-function}) is applied, the
following least-favorable distribution \cite{montanari2012graphical}
turns out to be{\small{}
\begin{equation}
p_{x}=\frac{\epsilon}{2}\Delta_{x=-\mu}+\left(1-\epsilon\right)\Delta_{x=0}+\frac{\epsilon}{2}\Delta_{x=\mu}\label{eq:least-favorite-distribution}
\end{equation}
}where $\Delta$ denotes the Dirac delta function, $\mu\in\left(-\infty,\infty\right)$,
and $\epsilon\in\left(0,1\right]$ is the normalised sparsity level.

Note that the state evolution theorem \cite[Theorem 1]{5695122} requires
a bounded moment of (\ref{eq:least-favorite-distribution}) which
implies $x\in\left(-\infty,\infty\right)$. The analysis first assumes
$m,\ n\rightarrow\infty$ for any fixed $\mu$, and then allows $\mu\rightarrow\infty$.
In \cite{montanari2012graphical}, the author uses $p_{x}=\frac{\epsilon}{2}\Delta_{x=-\infty}+\left(1-\epsilon\right)\Delta_{x=0}+\frac{\epsilon}{2}\Delta_{x=\infty}$
for notational brevity.

The optimal threshold that minimises the MSE when (\ref{eq:least-favorite-distribution})
is applied, is given by{\small{}
\begin{equation}
\lambda^{t}\coloneqq\left(\alpha^{\dagger}\right)^{t}\sigma_{e}^{t},\ \left(\alpha^{\dagger}\right)^{t}=\arg\min_{\alpha\in\mathbb{R}_{+}}M\left(\epsilon,\alpha,\mu,\sigma_{e}^{t}\right),\label{eq:optimal_alpha}
\end{equation}
}where{\small{}
\begin{align}
M\left(\epsilon,\alpha,\mu,\sigma_{e}^{t}\right) & \!=\!\epsilon\left(\alpha^{2}\!+\!1\right)\Phi\left(\!-\!\alpha\!+\!\frac{\mu}{\sigma_{e}^{t}}\right)\!-\!\epsilon\left(\alpha\!+\!\frac{\mu}{\sigma_{e}^{t}}\right)\phi\left(\alpha\!-\!\frac{\mu}{\sigma_{e}^{t}}\right)\nonumber \\
 & \!+\!\epsilon\frac{\mu^{2}}{\left(\sigma_{e}^{t}\right)^{2}}\left(\Phi\left(\alpha-\frac{\mu}{\sigma_{e}^{t}}\right)-\Phi\left(-\alpha-\frac{\mu}{\sigma_{e}^{t}}\right)\right)\nonumber \\
 & \!+\!\epsilon(\alpha^{2}\!+\!1)\Phi\left(\!-\alpha\!-\!\frac{\mu}{\sigma_{e}^{t}}\right)\!-\!\epsilon\left(\alpha\!-\!\frac{\mu}{\sigma_{e}^{t}}\right)\phi\left(\!-\alpha\!-\!\frac{\mu}{\sigma_{e}^{t}}\right)\nonumber \\
 & +(1-\epsilon)\left[2(\alpha^{2}+1)\Phi\left(-\alpha\right)-2\alpha\phi\left(\alpha\right)\right],\label{eq:M-function0}
\end{align}
}$\phi(x)$ is the standard Gaussian density and $\Phi(x)=\int_{-\infty}^{x}\phi(t)dt$
is the corresponding cumulative distribution function. The MSE at
each iteration is the{\small{}n
\begin{equation}
{\rm Err}_{t+1}=M\left(\epsilon,\left(\alpha^{\dagger}\right)^{t},\mu,\sigma_{e}^{t}\right)\left(\sigma_{e}^{t}\right)^{2}.\label{eq:MSE-least-favorite}
\end{equation}
}{\small\par}

Applying state evolution with these results and letting $\mu\rightarrow\infty$
leads to the following theorem.
\begin{thm}
\label{thm:Theorem1} For a linear measurement system (\ref{eq:linear system 1})
with signal model (\ref{eq:least-favorite-distribution}) ($\mu\rightarrow\infty$)
and additive white Gaussian noise with variance (\ref{eq:noise model}),
apply the AMP algorithm with estimator (\ref{eq:soft-threshold-function}).
The optimal number of measurements in an MSE sense is asymptotically
given b{\small{}y
\begin{equation}
\delta^{\dagger}=2M\left(\epsilon,\alpha^{\dagger}\right),\label{eq:optimal-delta}
\end{equation}
} which is independent of the noise variance.
\end{thm}
\begin{IEEEproof}
By the convergence assumption, when $t\rightarrow\infty$, we have
$\sigma_{e}^{t+1}=\sigma_{e}^{t}=\sigma_{e}^{\infty}$ and ${\rm Err}_{t+1}={\rm Err}_{t}={\rm Err}_{\infty}$.
Substituting (\ref{eq:SE_COMMON}) into (\ref{eq:MSE-least-favorite})
results in
\[
{\rm Err}_{\infty}=M\left(\epsilon,\left(\alpha^{\dagger}\right)^{\infty},\mu,\sigma_{e}^{\infty}\right)\left(\frac{1}{\delta}{\rm Err}_{\infty}+\delta\sigma_{0}^{2}\right).
\]
For the worst case $\mu\rightarrow\infty$, (\ref{eq:M-function0})
is independent of $\sigma_{e}^{t}$, and $\alpha^{\dagger}$ becomes
a constant. We then have
\begin{align}
 & M\left(\epsilon,\alpha^{\dagger}\right)\coloneqq\lim_{\mu\rightarrow\infty}M\left(\epsilon,\left(\alpha^{\dagger}\right)^{\infty},\mu,\sigma_{e}^{\infty}\right)\nonumber \\
 & \quad=\epsilon\left(1\!+\!\left(\alpha^{\dagger}\right)^{2}\right)\nonumber \\
 & \quad\quad+\!\left(1\!-\!\epsilon\right)\left[2\left(1\!+\!\left(\alpha^{\dagger}\right)^{2}\right)\Phi\left(\!-\alpha^{\dagger}\right)\!-\!2\alpha^{\dagger}\phi\left(\alpha^{\dagger}\right)\right]\label{eq:M-function}
\end{align}
and
\begin{eqnarray}
{\rm Err}_{\infty} & = & \frac{M\left(\epsilon,\alpha^{\dagger}\right)\delta^{2}\sigma_{0}^{2}}{\delta-M\left(\epsilon,\alpha^{\dagger}\right)}.\label{eq:Err_inf_function}
\end{eqnarray}
Consider ${\rm Err}_{\infty}$ as a function of $\delta$. Take the
derivative with respect to $\delta$, and equate it to zero. For $\delta>M\left(\epsilon,\alpha^{\dagger}\right)$
(which ensures that ${\rm Err}_{\infty}$ is a positive value), we
have a unique saddle point $\delta^{\dagger}=2M\left(\epsilon,\alpha^{\dagger}\right)$
(a local minima or a local maxima). As $\delta\rightarrow\infty$,
we have ${\rm Err}_{\infty}\rightarrow\infty$, thus, $\delta^{\dagger}$
is a local minima which is our required solution. In addition, $\delta^{\dagger}$
does not depend on the noise base level $\sigma_{0}^{2}$.
\end{IEEEproof}

\subsection{\label{subsec:Bernoulli-Gaussian-Distribution}Bernoulli-Gaussian
Distribution}

Next we consider the Bernoulli-Gaussian prior \cite{Lu2016_2,Lu2016,OnJointRecoveryXiaoChen}
with probability density given by
\begin{align}
p_{x} & =\left(1-\epsilon\right)\Delta_{x=0}+\epsilon p_{G}\left(x;\,0,\,\sigma_{x}^{2}\right),\label{eq:BG-distribution}
\end{align}
where {\small{}$p_{G}\left(x;\,0,\,\sigma_{x}^{2}\right)$} represents
the zero mean Gaussian density with variance $\sigma_{x}^{2}$.

The $\eta\left(\cdot\right)$ function in (\ref{eq:AMP1}) is designed
based on the prior information of $\bm{x}$. Le{\small{}t
\begin{equation}
R^{t}\coloneqq\frac{\sigma_{x}^{2}}{\left(\sigma_{e}^{t}\right)^{2}+\sigma_{x}^{2}}\label{eq:Rt}
\end{equation}
 }and define
\begin{align}
I\left(R^{t},\epsilon\right) & \coloneqq\int\frac{\phi\left(x\right)}{1+\frac{1-\epsilon}{\epsilon}\frac{1}{\sqrt{1-R^{t}}}{\rm exp}\left(-\frac{R^{t}}{1-R^{t}}\frac{x^{2}}{2}\right)}x^{2}dx.\label{eq:I1}
\end{align}
The element-wise function $\eta\left(\cdot\right)$ takes the mean
value of the posterior probability $p\left(x|\beta_{i}^{t}\right)$
which provides the MMSE estimate \cite{OnJointRecoveryXiaoChen}.
For each element of $\bm{\beta}^{t}$,
\begin{align}
\eta\left(x_{i}^{t}|\beta_{i}^{t}\right) & \coloneqq{\rm \mathbb{E}}\left[x|\beta_{i}^{t}\right]=\frac{p_{G}\left(\beta_{i}^{t};0,\left(\sigma_{e}^{t}\right)^{2}+\sigma_{x}^{2}\right)}{p\left(\beta_{i}^{t}\right)}\epsilon R^{t}\beta_{i}^{t},\label{eq:eta_BG_real}
\end{align}
{\small{}with
\[
p\left(\beta_{i}^{t}\right)\coloneqq\left(1-\epsilon\right)p_{G}\left(\beta_{i}^{t};0,\left(\sigma_{e}^{t}\right)^{2}\right)+\epsilon p_{G}\left(\beta_{i}^{t};0,\left(\sigma_{e}^{t}\right)^{2}+\sigma_{x}^{2}\right).
\]
}The corresponding derivative of $\eta\left(x_{i}^{t}|\beta_{i}^{t}\right)$
is calculated via
\[
\eta^{\prime}\left(x_{i}^{t}|\beta_{i}^{t}\right)=\frac{R^{t}}{v_{3}+1}+\frac{R^{t}v_{3}v_{2}\left(\beta_{i}^{t}\right)^{2}}{\left(v_{3}+1\right)^{2}}
\]
wher{\small{}e
\begin{align*}
v_{1}\!\coloneqq\!\frac{1-\epsilon}{\epsilon}\!\sqrt{\!\frac{\left(\sigma_{e}^{t}\right)^{2}\!+\!\sigma_{x}^{2}}{\left(\sigma_{e}^{t}\right)^{2}}},\  & v_{2}\!\coloneqq\!\frac{R^{t}}{\left(\sigma_{e}^{t}\right)^{2}},\ v_{3}\!\coloneqq\!v_{1}\exp\left(\!-\!\frac{1}{2}v_{2}\left(\beta_{i}^{t}\right)^{2}\right),
\end{align*}
}and the MSE is given by
\begin{equation}
{\rm Err}_{t+1}\coloneqq\left[\frac{R^{t}\epsilon}{1-R^{t}}\left(1-R^{t}I\left(R^{t},\epsilon\right)\right)\right]\left(\sigma_{e}^{t}\right)^{2}.\label{eq:Err_BG}
\end{equation}

\begin{lem}
\label{lem:Lemma1}\cite[Lemma 2]{GAMP} Consider a random variable
U with conditional probability density function of the form $p_{U|V}\left(u|v\right)\coloneqq\frac{1}{Z\left(v\right)}exp\left(\phi_{0}\left(u\right)+uv\right),$
where $Z\left(v\right)$ is a normalization constant. Then,
\begin{align*}
\frac{\partial}{\partial v}{\rm log}Z\left(v\right) & ={\rm E}\left[U|V=v\right]\\
\frac{\partial^{2}}{\partial v^{2}}{\rm log}Z\left(v\right) & =\frac{\partial}{\partial v}{\rm E}\left[U|V=v\right]={\rm var}\left(U|V=v\right).
\end{align*}
\end{lem}
Based on Lemma \ref{lem:Lemma1} above, ${\rm Err}_{t+1}$ can be
approximated as
\begin{equation}
{\rm Err}_{t+1}\approx\left[\frac{1}{n}\sum_{i=1}^{n}\eta^{\prime}\left(x_{i}^{t}|\beta_{i}^{t}\right)\right]\left(\sigma_{e}^{t}\right)^{2},\label{eq:FastCalculation1}
\end{equation}
which avoids the integration of $I\left(R^{t},\epsilon\right)$ in
(\ref{eq:Err_BG}). Equation (\ref{eq:Err_BG}) is used in the SE
analysis while (\ref{eq:FastCalculation1}) should be used in signal
reconstruction.

\begin{proof}[Proof of (\ref{eq:FastCalculation1})]

Now go back to (\ref{eq:equivalent signal}) and define $\bm{B}(t)=\left(\bm{A}\left(t\right)^{*}\bm{A}\left(t\right)-\bm{I}\right)$.
We borrow the statements from \cite[Section C]{5695122} which claimed
that, based on the central limit theorem, each entry of $\bm{B}(t)$
is approximately normal with zero mean and variance $\frac{1}{m}$
and $\bm{B}(t)\left(\bm{x}-\bm{x}^{t}\right)$ converges to a vector
with i.i.d. normal entries. In addition, according to the law of large
numbers, $\bm{A}\left(t\right)^{*}\bm{w}$ is also a vector of i.i.d.
normal entries with mean zero and variance that converges to $\sigma_{w}^{2}$,
which is approximately independent of $\bm{B}(t)\left(\bm{x}-\bm{x}^{t}\right)$.
Thus each entry of $\bm{\beta}^{t}$ converges to $x+w_{e}^{t}$ where
$x\sim p_{x}$ and $w_{e}^{t}\sim\mathcal{N}(0,\left(\sigma_{e}^{2}\right)^{t})$.
Consider the conditional probability
\begin{align*}
p\left(x|\beta\right)=\frac{p\left(x,\beta\right)}{p\left(\beta\right)}= & \frac{\left(1-\epsilon\right)p_{G}\left(\beta-x;0,\sigma_{e}^{2}\right)\Delta_{x=0}}{p\left(\beta\right)}\\
 & +\frac{\epsilon p_{G}\left(x;0,\sigma_{x}^{2}\right)p_{G}\left(\beta-x;0,\sigma_{e}^{2}\right)}{p\left(\beta\right)},
\end{align*}
in which we only care about the second term (the first term has no
contribution to $\mathbb{E}\left[x|\beta\right]$ and ${\rm var}\left[x|\beta\right]$
due to $\Delta_{x=0}$). The numerator of the second term is
\begin{align*}
 & \epsilon p_{G}\left(x;0,\sigma_{x}^{2}\right)p_{G}\left(\beta-x;0,\sigma_{e}^{2}\right)\\
 & =\epsilon\frac{1}{\sqrt{2\pi\sigma_{x}^{2}}}{\rm exp}\left(-\frac{x^{2}}{2\sigma_{x}^{2}}\right)\frac{1}{\sqrt{2\pi\sigma_{e}^{2}}}{\rm exp}\left(-\frac{\left(\beta-x\right)^{2}}{2\sigma_{e}^{2}}\right)\\
 & =\epsilon\frac{1}{2\pi\sigma_{x}\sigma_{e}}{\rm exp}\left(-\frac{\beta^{2}}{2\sigma_{e}^{2}}\right){\rm exp}\left(\frac{-\sigma_{e}^{2}-\sigma_{x}^{2}}{2\sigma_{x}^{2}\sigma_{e}^{2}}x^{2}+\frac{x\beta}{\sigma_{e}^{2}}\right).
\end{align*}
Dividing the numerator and denominator of $\frac{1}{p\left(\beta\right)}\epsilon p_{G}\left(x;0,\sigma_{x}^{2}\right)p_{G}\left(\beta-x;0,\sigma_{e}^{2}\right)$
by $\epsilon\frac{1}{2\pi\sigma_{x}\sigma_{e}}{\rm exp}\left(-\frac{\beta^{2}}{2\sigma_{e}^{2}}\right)$,
the remaining part in the numerator is ${\rm exp}\left(\frac{-\sigma_{e}^{2}-\sigma_{x}^{2}}{2\sigma_{x}^{2}\sigma_{e}^{2}}x^{2}+\frac{x\beta}{\sigma_{e}^{2}}\right)$.
Comparing the term ${\rm exp}\left(\frac{-\sigma_{e}^{2}-\sigma_{x}^{2}}{2\sigma_{x}^{2}\sigma_{e}^{2}}x^{2}+\frac{x\beta}{\sigma_{e}^{2}}\right)$
with the one ${\rm exp}\left(\phi_{0}\left(u\right)+uv\right)$ in
Lemma \ref{lem:Lemma1}, we have $u=\frac{x}{\sigma_{e}^{2}}$ and
$v=\beta$. Based on (\ref{eq:eta_BG_real}) and Lemma \ref{lem:Lemma1},
we get
\begin{align*}
\mathbb{E}\left[U|V=\beta\right] & =\frac{\mathbb{E}\left[X|V=\beta\right]}{\sigma_{e}^{2}}=\frac{\eta\left(\beta\right)}{\sigma_{e}^{2}},\\
{\rm var}\left(U|V=\beta\right)\! & \!=\!\frac{\eta^{\prime}\left(\beta\right)}{\sigma_{e}^{2}}\!=\!{\rm var}\left(\!\frac{X}{\sigma_{e}^{2}}|V\!=\!\beta\!\right)\!=\!\frac{1}{\sigma_{e}^{4}}{\rm var}\left(X|V\!=\!\beta\right),
\end{align*}
so that
\[
{\rm var}\left(X|V=\beta\right)=\eta^{\prime}\left(\beta\right)\sigma_{e}^{2}.
\]
Since the MSE is the average of ${\rm var}\left(X|V=\beta\right)$
with respect to different $\beta$'s, (\ref{eq:FastCalculation1})
follows. \end{proof}

In the Bernoulli-Gaussian case, there are no closed-form representations
of $\sigma_{e}^{\infty}$ and ${\rm Err}_{\infty}$. These two values
only can be obtained numerically. The following shows a general process
of finding the optimal value of $\delta$ to minimise the recovery
MSE (${\rm Err}_{\infty}$).

When AMP converges, $\sigma_{e}^{t+1}=\sigma_{e}^{t}=\sigma_{e}^{\infty}$
and ${\rm Err}_{t+1}={\rm Err}_{t}={\rm Err}_{\infty}$. Based on
(\ref{eq:SE_COMMON}), we have
\begin{align}
\left(\sigma_{e}^{\infty}\right)^{2} & =\frac{1}{\delta}{\rm Err}_{\infty}+\delta\sigma_{0}^{2},\label{eq:ConvergesMustHoldEquation}
\end{align}
where ${\rm Err}_{\infty}$ is a function of $\epsilon,\ \sigma_{x}^{2},\ \sigma_{0}^{2}$
and $\sigma_{e}^{\infty}$. With fixed $\epsilon,\ \sigma_{x}^{2}$
and $\sigma_{0}^{2}$, ${\rm Err}_{\infty}$ is a function of $\left(\sigma_{e}^{\infty}\right)^{2}$
and vice versa. For any given ${\rm Err}_{\infty}$, (\ref{eq:ConvergesMustHoldEquation})
is a quadratic equation of $\delta$. The values of $\delta$ that
achieve ${\rm Err}_{\infty}$ are given by
\begin{align}
\delta & =\frac{\left(\sigma_{e}^{\infty}\right)^{2}\pm\sqrt{\left(\sigma_{e}^{\infty}\right)^{4}-4\sigma_{0}^{2}{\rm Err}_{\infty}}}{2\sigma_{0}^{2}}.\label{eq:required_delta}
\end{align}
According to our numerical results, if ${\rm Err}_{\infty}$ is too
small, then there is no valid $\delta$ (must be a real value). This
means that such ${\rm Err}_{\infty}$ is not achievable no matter
how we design $\delta$. Increase ${\rm Err}_{\infty}$ until
\begin{equation}
\left(\sigma_{e}^{\infty}\right)^{4}=4\sigma_{0}^{2}{\rm Err}_{\infty}\label{eq:critical condition}
\end{equation}
which provides a unique optimal solution
\begin{equation}
\delta^{\dagger}=\frac{(\sigma_{e}^{\infty})^{2}}{2\sigma_{0}^{2}}.\label{eq:optimal delta general}
\end{equation}
The corresponding ${\rm Err}_{\infty}$ is the minimum value that
is achievable.

The conclusions achieved above can be used to derive the result of
Theorem \ref{thm:Theorem1}. Recall that in the worst case analysis,
based on (\ref{eq:MSE-least-favorite}), we have ${\rm Err}_{\infty}=M\left(\epsilon,\alpha^{\dagger}\right)\left(\sigma_{e}^{\infty}\right)^{2}$.
Substituting this ${\rm Err}_{\infty}$ into (\ref{eq:critical condition})
provides $\left(\sigma_{e}^{\infty}\right)^{2}=4\sigma_{0}^{2}M\left(\epsilon,\alpha^{\dagger}\right)$.
Further substituting this $\left(\sigma_{e}^{\infty}\right)^{2}$
into (\ref{eq:optimal delta general}) results in $\delta^{\dagger}=\frac{4\sigma_{0}^{2}M\left(\epsilon,\alpha^{\dagger}\right)}{2\sigma_{0}^{2}}=2M\left(\epsilon,\alpha^{\dagger}\right)$
which coincides with the solution of Theorem \ref{thm:Theorem1}.

Note that it is proved in \cite{1742-5468-2012-08-P08009} that for
a region of parameters $(\epsilon,\ \delta,\ \sigma_{e}^{\infty},\ \sigma_{x}^{2},\ \sigma_{0}^{2})$,
belief propagation based algorithms (i.e. AMP) may provide a suboptimal
solution compared with the one achieved by optimal Bayesian inference
(the best possible reconstruction, regardless of the algorithms).
In \cite{1742-5468-2012-08-P08009}, it also shows that the suboptimal
solution provided by AMP will converge to the optimal solution when
the noise variance grows. In this paper, we focus on AMP reconstruction
only and do not consider the best possible reconstruction provided
by other algorithms.

\subsection{\label{subsec:Non-sparse-Case-(Gaussian)}Special Case when $\epsilon=1$
(Gaussian)}

Consider the Bernoulli-Gaussian prior with $\epsilon=1$. In this
case, (\ref{eq:I1}) degenerates to the variance of a standard Gaussian
distribution which is a constant with value equal to $1$. The estimated
error (\ref{eq:Err_BG}) is then
\begin{equation}
{\rm Err}_{t+1}=R^{t}\left(\sigma_{e}^{t}\right)^{2},\label{eq:Err_Gaussian}
\end{equation}
where $R^{t}$ is given by (\ref{eq:Rt}). Substituting (\ref{eq:Err_Gaussian})
and (\ref{eq:noise model}) into (\ref{eq:SE_COMMON}) and setting
$t=t+1=\infty$ yields
\begin{align}
\left(\sigma_{e}^{\infty}\right)^{2} & =\frac{1}{\delta}{\rm Err}_{\infty}+\delta\sigma_{0}^{2}\label{eq:Gaussian_key1}\\
 & =\frac{1}{\delta}\frac{\sigma_{x}^{2}\left(\sigma_{e}^{\infty}\right)^{2}}{\left(\sigma_{e}^{\infty}\right)^{2}+\sigma_{x}^{2}}+\delta\sigma_{0}^{2}.\label{eq:Gaussian_key2}
\end{align}
Based on (\ref{eq:Gaussian_key2}), we have
\begin{equation}
\left(\sigma_{e}^{\infty}\right)^{2}=\frac{\left(c\sigma_{x}^{2}+\delta\sigma_{0}^{2}\right)+\sqrt{\left(c\sigma_{x}^{2}+\delta\sigma_{0}^{2}\right)^{2}+4\sigma_{x}^{2}\delta\sigma_{0}^{2}}}{2},\label{eq:Gaussian_key3}
\end{equation}
where $c\coloneqq\frac{\left(1-\delta\right)}{\delta}$. We ignore
the negative value due to the non-negative property of the error.
Using (\ref{eq:Gaussian_key1}), the final estimation error at the
fixed point is
\begin{align}
{\rm Err}_{\infty} & =\delta\left(\left(\sigma_{e}^{\infty}\right)^{2}-\delta\sigma_{0}^{2}\right).\label{eq:Gaussian_key4}
\end{align}
Substituting (\ref{eq:Gaussian_key3}) into (\ref{eq:Gaussian_key4}),
we obtain (\ref{eq:RandomMatrixTheory_Gaussian}).

\section{\label{sec:Analysis-in-Complex}Analysis in Complex Domain}

We now consider the case in which all the elements of $\bm{y}$, $\bm{x}$,
$\bm{A}$ and $\bm{w}$ in (\ref{eq:linear system 1}) are complex
values. The analysis in the complex domain follows the same line as
in the real setting but with different formulas for $M\left(\epsilon,\alpha\right)$
and ${\rm Err}_{t}$. 

\textbf{Least-favorable distribution}: The complex AMP (CAMP) algorithm
for a least-favorable distribution has been analysed in \cite{6478821}
with a new Onsager term. Based on \cite{6478821}, the $\eta$ {\small{}function
is
\begin{equation}
\eta\left(\beta_{i}^{t},\lambda^{t}\right)\coloneqq\left(\beta_{i}^{t}-\frac{\lambda^{t}\left(\beta_{i}^{t}\right)}{\left|\beta_{i}^{t}\right|}\right)\mathrm{{\rm \bm{1}}}_{\left\{ \left|\beta_{i}^{t}\right|>\lambda^{t}\right\} }\label{eq:eta_soft_complex}
\end{equation}
}where {\small{}$\mathrm{{\rm \bm{1}}}_{\left\{ \left|\beta_{i}^{t}\right|>\lambda^{t}\right\} }$}
denotes the indicator function. The least-favorable distribution becomes
$p_{\left|x\right|}=\left(1-\epsilon\right)\Delta_{\left|x\right|=0}+\epsilon\Delta_{\left|x\right|=+\infty}$
with the assumption that the phase of $x$ is isotropic.

The formula of ${\rm Err}_{C,t}$ is the same as in real case but
with a new function $M_{C}\left(\epsilon,\alpha\right)${\footnotesize{}:
\begin{equation}
M_{C}\left(\epsilon,\alpha\right)\!\coloneqq\!\epsilon\left(1\!+\!\alpha^{2}\right)\!+\!\left(1\!-\!\epsilon\right)\!\left[\sqrt{2\pi}\phi\left(\sqrt{2}\alpha\right)\!-\!2\alpha\sqrt{\pi}\Phi\left(\!-\!\sqrt{2}\alpha\right)\right].\label{eq:Mc-fuction}
\end{equation}
}Comparing (\ref{eq:Mc-fuction}) with (\ref{eq:M-function}), we
see that the estimation error of the non-zero components of the signal
are the same (first term). The difference between them stems from
the de-noising for the zero components of the signal (second term).
For a complete derivation of the new Onsager term and calculation
of $\eta^{\prime}\left(\beta_{i}^{t},\lambda\right)$, we refer the
reader to \cite{6478821}. 

\textbf{Bernoulli-Gaussian distribution}: We assume that the real
part and imaginary part of the complex variable share the same mean
and variance and their magnitudes are uncorrelated. Let {\small{}$x\sim\mathcal{CN}\left(\mu,\sigma_{x}^{2}\right)$}.
Then we have {\small{}$\left(x\right)^{R},\left(x\right)^{I}\sim\mathcal{N}\left(\mu,\frac{\sigma_{x}^{2}}{2}\right)$.}
Under this assumption{\small{},
\begin{align}
p_{CG}\left(x;\mu,\sigma_{x}^{2}\right) & =p_{G}\left(\left(x\right)^{R};\mu,\frac{\sigma_{x}^{2}}{2}\right)p_{G}\left(\left(x\right)^{I};\mu,\frac{\sigma_{x}^{2}}{2}\right)\nonumber \\
 & =\frac{1}{\pi\sigma_{x}^{2}}{\rm exp}\left(-\frac{\left|x-\mu_{c}\right|^{2}}{\sigma_{x}^{2}}\right),\label{eq:complex Gaussian}
\end{align}
}where $\mu_{c}=\mu+i\mu$ and the Bernoulli-Gaussian distribution
in the complex domain becomes $p(x)=\left(1-\epsilon\right)\Delta_{\left|x\right|=0}+\epsilon p_{CG}\left(x\right)$.
For the estimator $\eta$, we just replace the $p_{G}$ probability
in (\ref{eq:eta_BG_real}) with $p_{CG}$ defined above. 

Now let
\[
\begin{array}{c}
p_{\beta,1}^{t}\!\coloneqq\!p_{CG}\left(\beta_{j}^{t};0,\left(\sigma_{e}^{t}\right)^{2}\!+\!\sigma_{x}^{2}\right),\,p_{\beta,2}^{t}\!\coloneqq\!p_{CG}\left(\beta_{j}^{t};0,\left(\sigma_{e}^{t}\right)^{2}\right),\\
p_{\beta,3}^{t}\coloneqq\left(1-\epsilon\right)p_{\beta,2}^{t}+\epsilon p_{\beta,1}^{t},
\end{array}
\]
 and
\[
p_{o}^{t}\coloneqq-\frac{2}{\sigma_{x}^{2}+\left(\sigma_{e}^{t}\right)^{2}}p_{\beta,3}^{t}+\frac{2\left(1-\epsilon\right)}{\sigma_{w}^{2}}p_{\beta,2}^{t}+\frac{2\epsilon}{\sigma_{x}^{2}+\left(\sigma_{e}^{t}\right)^{2}}p_{\beta,1}^{t}.
\]
The four derivatives of $\eta$ can be calculated based on the following
formulas{\small{}:
\begin{align}
\frac{\partial\eta^{R}\left(\beta_{j}^{t}\right)}{\partial\left(\beta_{j}^{t}\right)^{R}} & =\frac{p_{o}^{t}}{\left(p_{\beta,3}^{t}\right)^{2}}p_{\beta,1}^{t}\epsilon R\left(\left(\beta_{j}^{t}\right)^{R}\right)^{2}+\frac{p_{\beta,1}^{t}}{p_{\beta,3}^{t}}\epsilon R\\
\frac{\partial\eta^{R}\left(\beta_{j}^{t}\right)}{\partial\left(\beta_{j}^{t}\right)^{I}} & =\frac{\partial\eta^{I}\left(\beta_{j}^{t}\right)}{\partial\left(\beta_{j}^{t}\right)^{R}}=\frac{p_{o}^{t}}{\left(p_{\beta,3}^{t}\right)^{2}}p_{\beta,1}^{t}\epsilon R\left(\beta_{j}^{t}\right)^{R}\left(\beta_{j}^{t}\right)^{I}\\
\frac{\partial\eta^{I}\left(\beta_{j}^{t}\right)}{\partial\left(\beta_{j}^{t}\right)^{I}} & =\frac{p_{o}^{t}}{\left(p_{\beta,3}^{t}\right)^{2}}p_{\beta,1}^{t}\epsilon R\left(\left(\beta_{j}^{t}\right)^{I}\right)^{2}+\frac{p_{\beta,1}^{t}}{p_{\beta,3}^{t}}\epsilon R.
\end{align}
}Finally, (\ref{eq:Err_BG}) is replaced by{\footnotesize{} 
\begin{align}
{\rm Err}_{C,t+1} & =\left[\frac{R^{t}\epsilon}{1-R^{t}}\left(1-R^{t}I_{C}\left(R^{t},\epsilon\right)\right)\right]\left(\sigma_{e}^{t}\right)^{2},\label{eq:Err_BG_c}\\
I_{C}\!\left(\!R^{t},\!\epsilon\!\right)\!\! & =\!\!\int\!_{x^{R}}\!\int\!_{x^{I}}\!\frac{\phi_{C}\left(x\right)}{1\!+\!\frac{1\!-\!\epsilon}{\epsilon}\frac{1}{1\!-\!R^{t}}\!{\rm exp}\!\left(\!-\!\frac{R^{t}}{1\!-\!R^{t}}\!\left|x\right|^{2}\!\right)}\left|x\right|\!^{2}\!dx^{I}\!dx^{R},\label{eq:2d_integration}
\end{align}
}where $\phi_{C}\left(x\right)=p_{CG}\left(x;0,1\right)$ is the standard
complex normal distribution. 

As in the real domain, we can efficiently calculate (\ref{eq:Err_BG_c})
by focusing on the real part of the signal only{\small{},
\begin{align}
{\rm Err}_{C,t+1}^{R} & \approx\frac{1}{n}\sum_{j=1}^{n}\left(\frac{\partial\eta^{R}\left(\beta_{j}^{t}\right)}{\partial\left(\beta_{j}^{t}\right)^{R}}\right)\frac{\left(\sigma_{e}^{t}\right)^{2}}{2}\\
{\rm Err}_{C,t+1} & \approx2{\rm Err}_{C,t}^{R}=\frac{1}{n}\sum_{j=1}^{n}\left(\frac{\partial\eta^{R}\left(\beta_{j}^{t}\right)}{\partial\left(\beta_{j}^{t}\right)^{R}}\right)\left(\sigma_{e}^{t}\right)^{2}.
\end{align}
}Based on the assumption that the real and imaginary parts of the
complex random variable are i.i.d., the total MSE is twice the MSE
of the real part.

The optimal $\delta^{\dagger}$ can be determined by using the Theorems
in Section \ref{sec:Analysis-in-Real}, and replacing $M(\epsilon,\alpha^{\dagger})$
and ${\rm Err}_{t}$ with $M_{C}(\epsilon,\alpha^{\dagger})$ in (\ref{eq:Mc-fuction})
and ${\rm Err}_{C,t}$ in (\ref{eq:Err_BG_c}), respectively. 

\section{\label{sec:BoundsANDSimulation}Bounds and Simulation }

\subsection{\label{sec:Boundary-Analysis}Bounds Analysis}

\begin{figure}
\centering{}\includegraphics[scale=0.185]{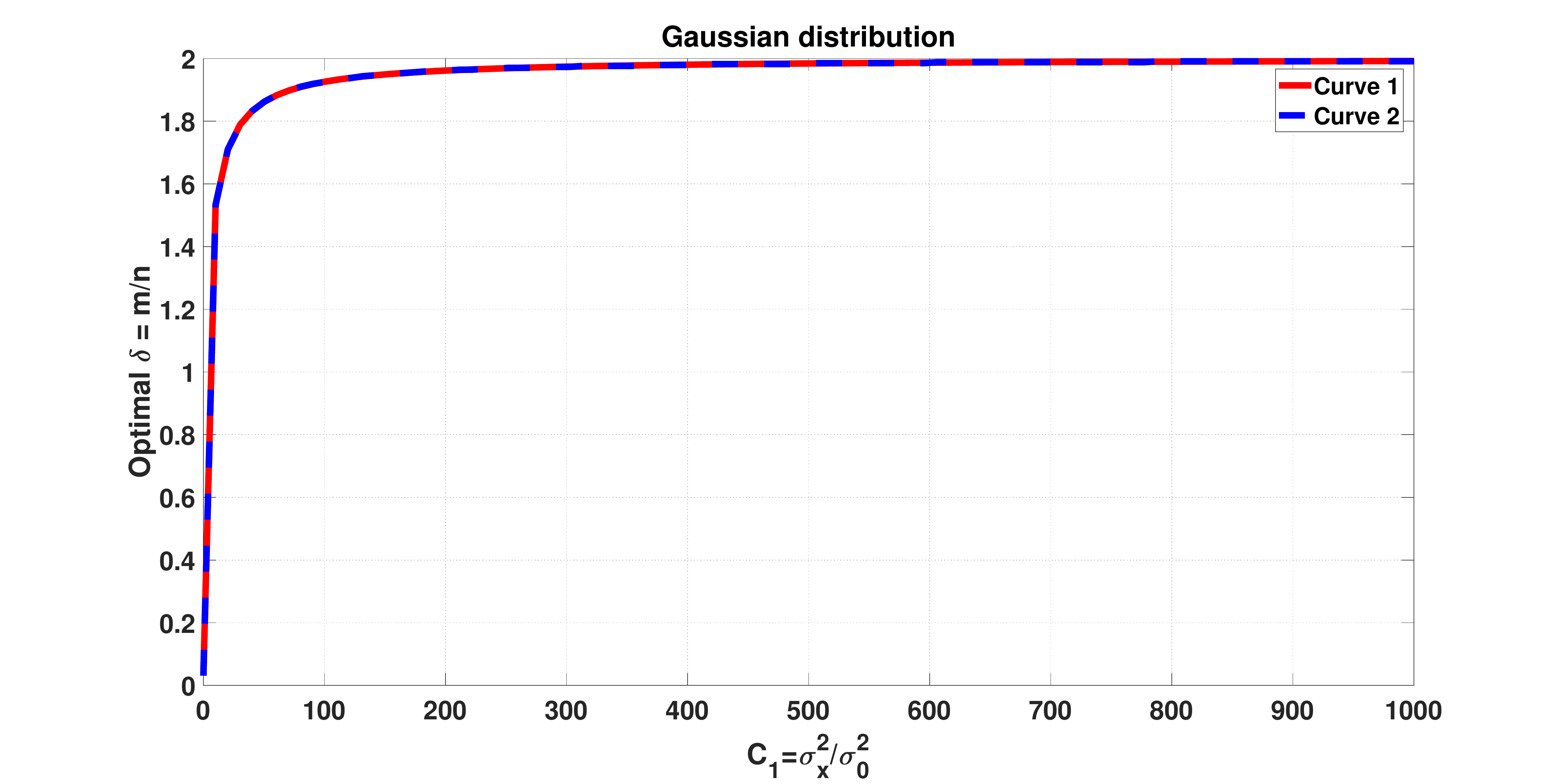}\caption{\label{fig:Gaussian-Case:-Optimal}Gaussian case: Optimal $\delta$
vs $C_{1}=\sigma_{x}^{2}/\sigma_{0}^{2}$. The red solid line (Curve
1) is directly achieved via (\ref{eq:optimal delta Gaussian}) and
the blue dashed line (Curve 2) is achieved via ${\rm argmin_{\delta}}{\rm Err}$
where ${\rm Err}$ is defined by (\ref{eq:RandomMatrixTheory_Gaussian}). }
\end{figure}
The optimal $\delta^{\dagger}$ for the least-favorable distribution
is given by (\ref{eq:optimal-delta}). For a given $\epsilon$, we
have a unique value of $M\left(\epsilon,\alpha^{\dagger}\right)$
to quantify $\delta^{\dagger}$. For the Bernoulli-Gaussian case,
we need to try different values of $\sigma_{e}^{\infty}$ to satisfy
the condition $\left(\sigma_{e}^{\infty}\right)^{4}=4\sigma_{0}^{2}{\rm Err}_{\infty}$,
thus $\delta^{\dagger}$ is computed numerically. In order to obtain
intuition into the values of $\delta^{\dagger}$ for different signals,
we derive bounds on $\delta^{\dagger}$ for the Gaussian, Bernoulli-Gaussian
and least-favorable distributions.
\begin{figure*}
\centering{}\subfloat[Real case]{\begin{centering}
\includegraphics[scale=0.18]{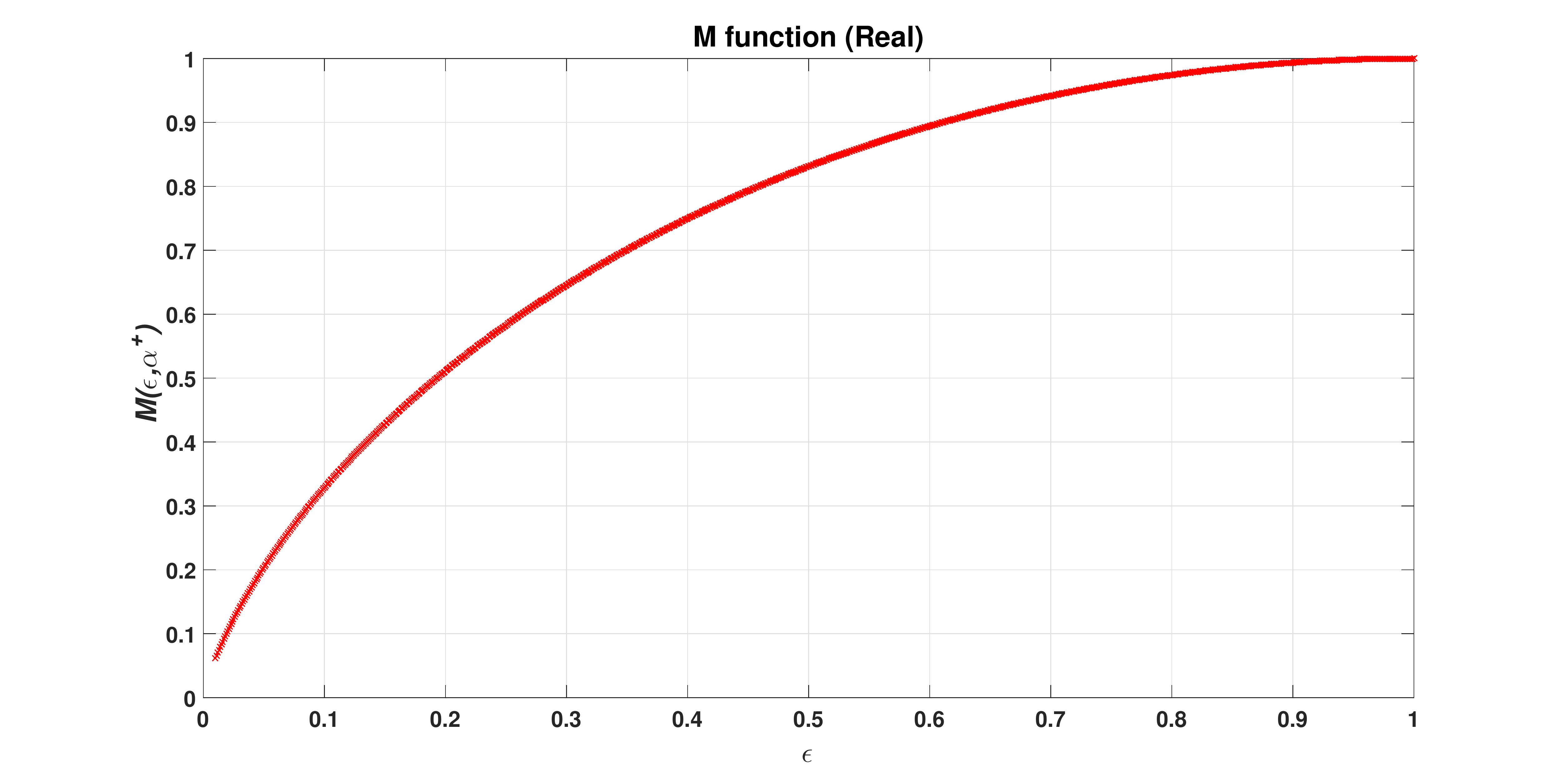}
\par\end{centering}
}\subfloat[Complex case]{\begin{centering}
\includegraphics[scale=0.18]{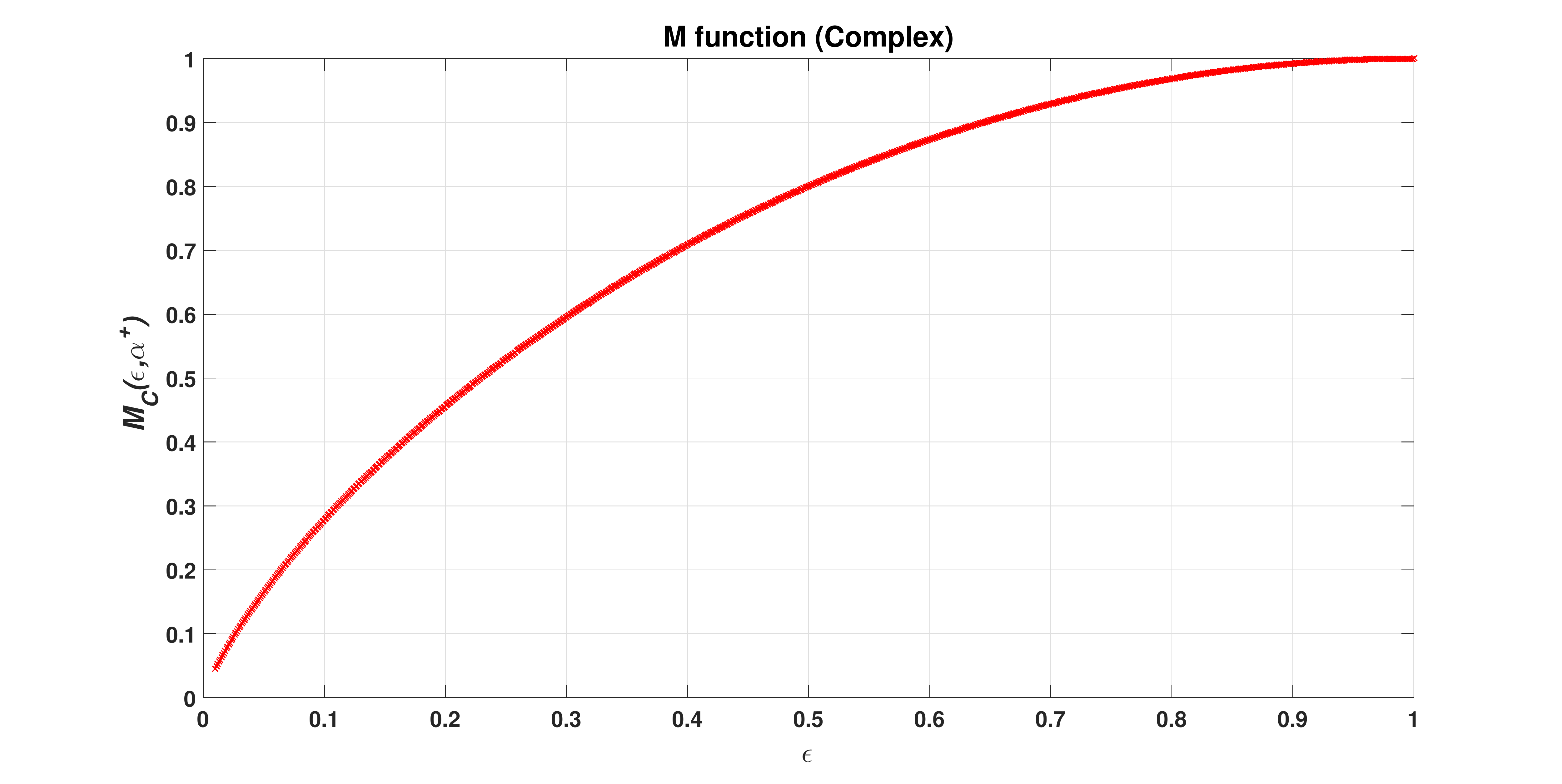}
\par\end{centering}
}\caption{\label{fig:M vs epsilon}$M\left(\epsilon,\alpha^{\dagger}\right)$
vs $\epsilon$.}
\end{figure*}

\begin{figure*}
\begin{centering}
\subfloat[3D plot (real case)]{\includegraphics[scale=0.18]{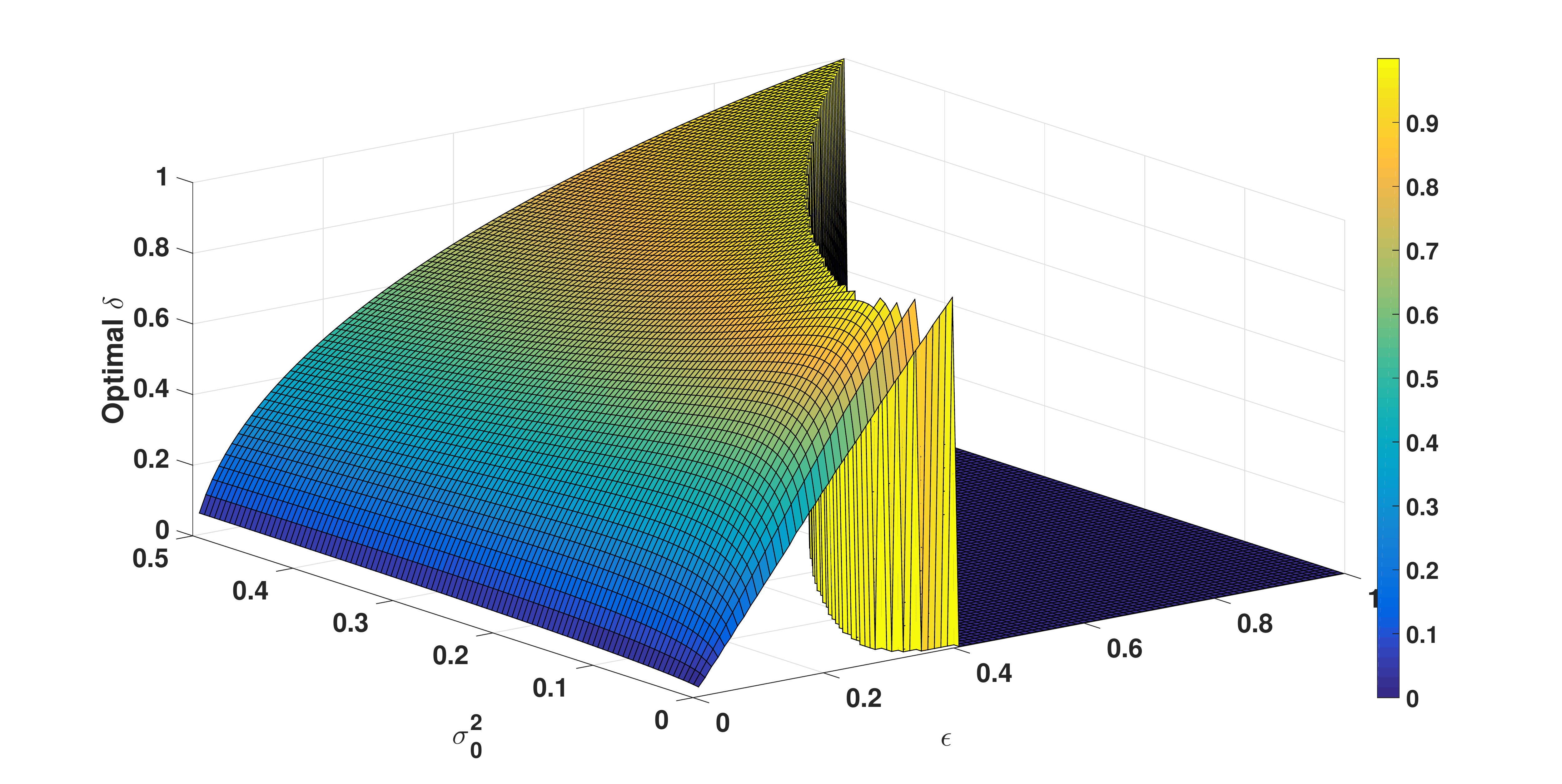}

}\subfloat[2D plot (real case)]{\includegraphics[scale=0.18]{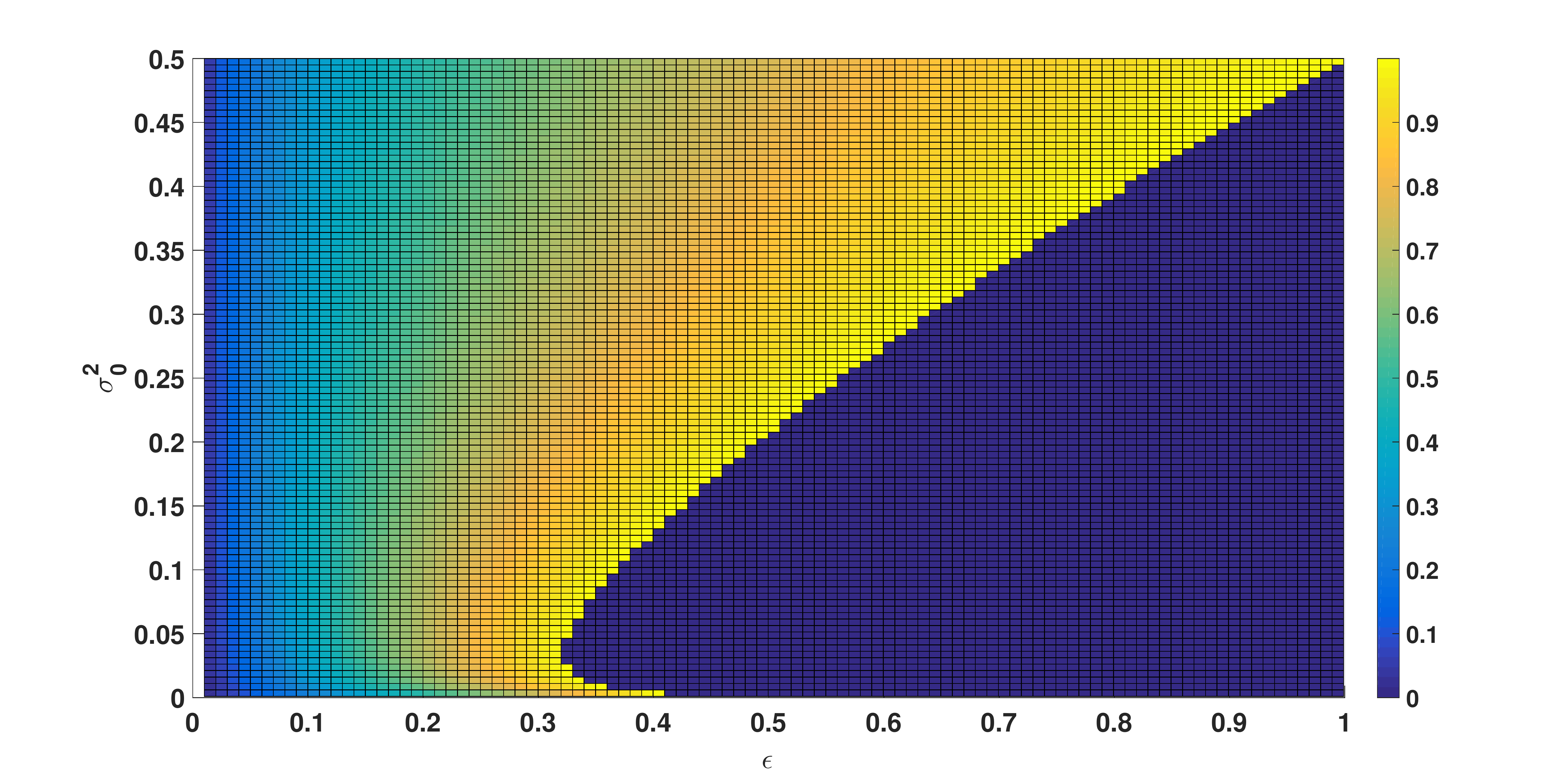}

}\\
\subfloat[3D plot (complex case)]{\includegraphics[scale=0.18]{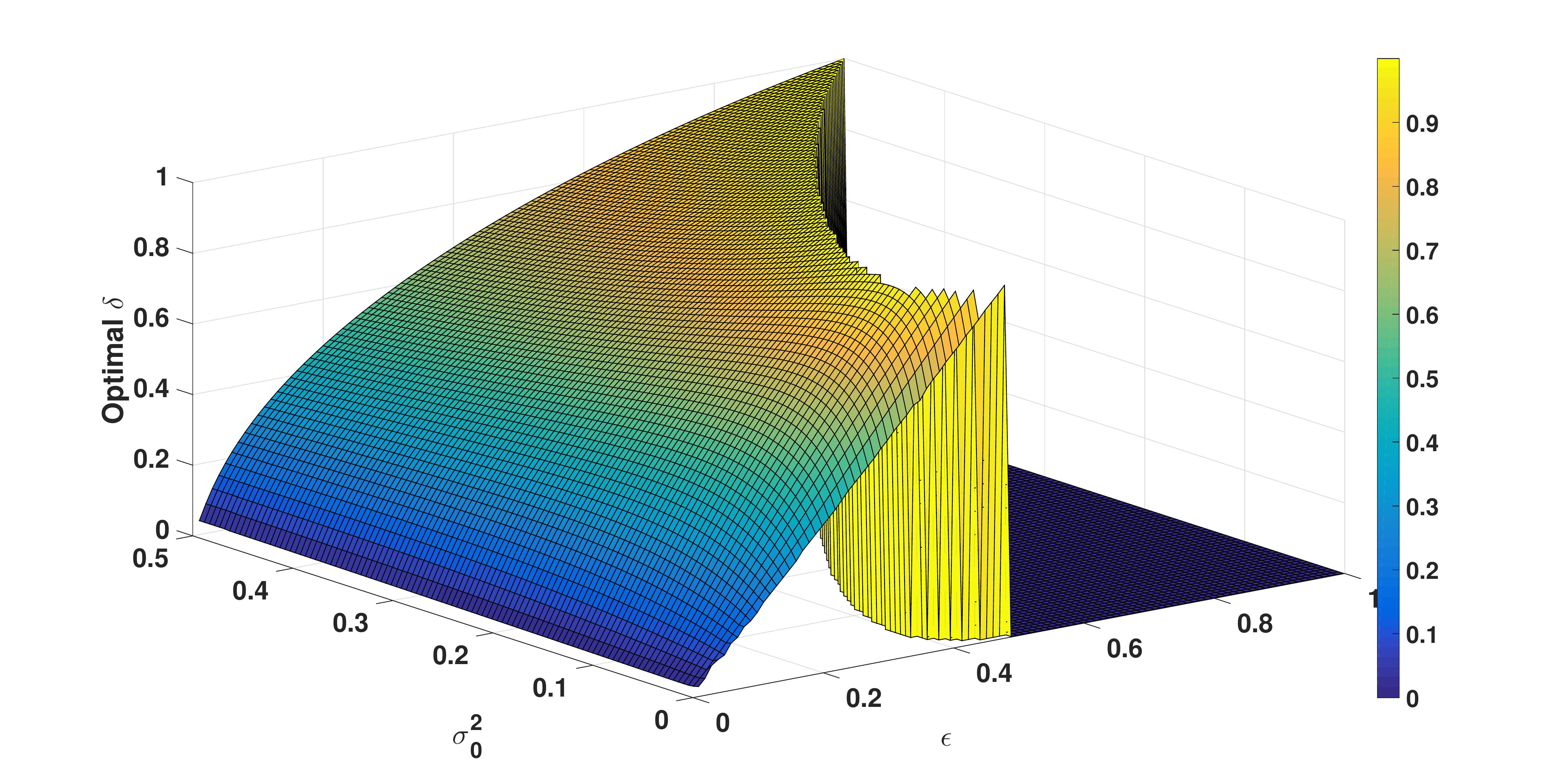}

}\subfloat[2D plot (complex case)]{\includegraphics[scale=0.18]{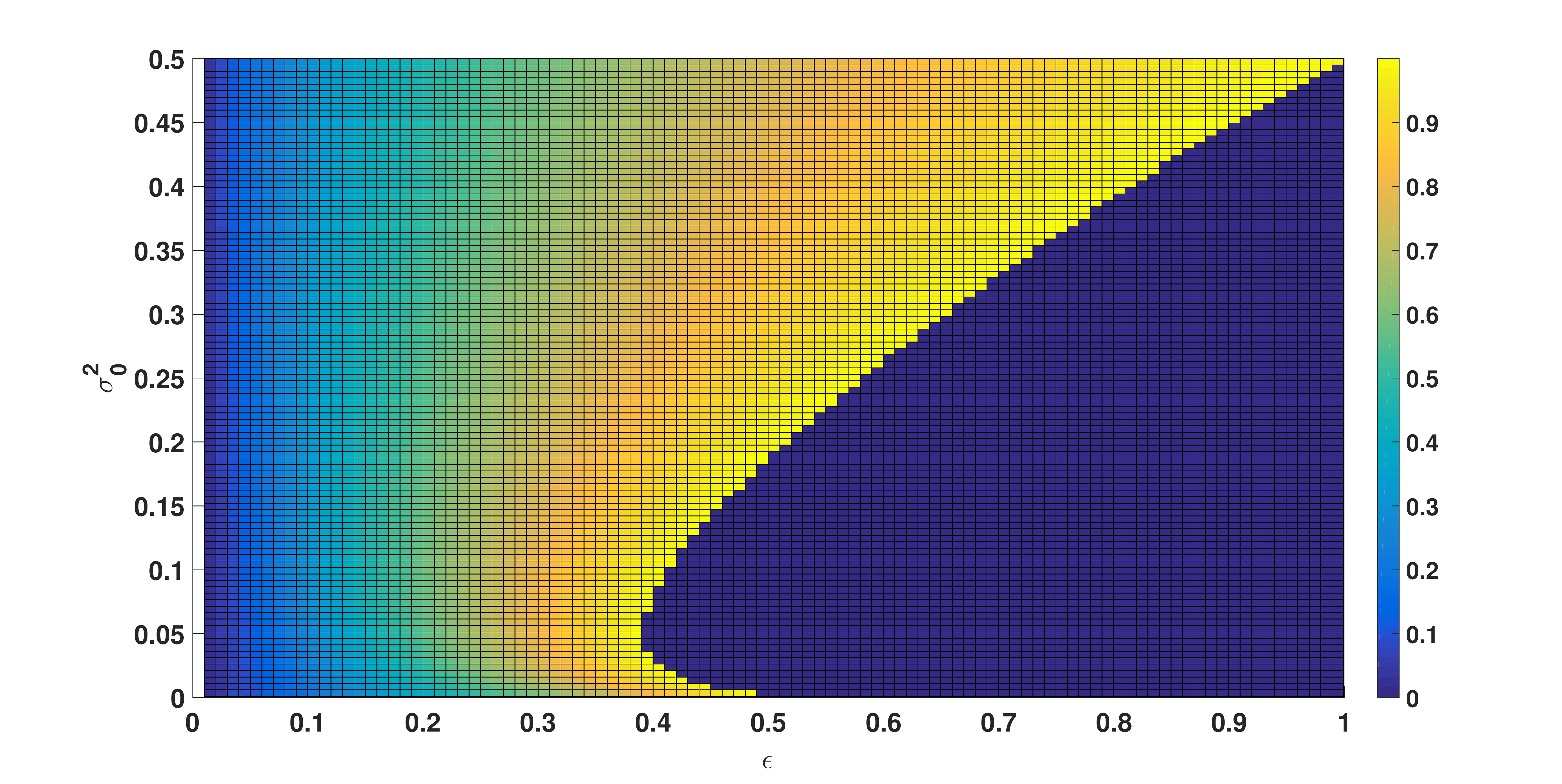}

}
\par\end{centering}
\centering{}\caption{\label{fig:Optimal_BG}Simulation of $\delta^{\dagger}$ with respect
to $\sigma_{0}^{2}$ and $\epsilon$ with fixed value $\sigma_{x}^{2}=1$.
Only the specific region $\delta^{\dagger}<1$ is plotted.}
\end{figure*}

\begin{prop}[Gaussian Distribution]
\label{claim:Gaussian}For a linear measurement system (\ref{eq:linear system 1})
with our proposed Gaussian noise model (\ref{eq:noise model}), assume
the signal elements are i.i.d. drawn from the Gaussian distribution
as in Section \ref{subsec:Non-Sparse-Setting} for both real and complex
cases. Then $\delta^{\dagger}<2$. In addition, when $\sigma_{x}^{2}<2\sigma_{0}^{2}$,
we have $\delta^{\dagger}<1$.
\end{prop}
\begin{IEEEproof}
We focus on (\ref{eq:RandomMatrixTheory_Gaussian}) and replace $\sigma_{w}^{2}$
with $\delta\sigma_{0}^{2}$. Let $g(\delta)=\frac{f(\delta)}{\sigma_{0}^{2}}$
where
\begin{align*}
f(\delta) & \coloneqq\left(-\delta^{2}\sigma_{0}^{2}+\left(1-\delta\right)\sigma_{x}^{2}\right)\\
 & +\sqrt{\left(\delta^{2}\sigma_{0}^{2}+\left(1-\delta\right)\sigma_{x}^{2}\right)^{2}+4\delta^{3}\sigma_{0}^{2}\sigma_{x}^{2}}.
\end{align*}
Then
\begin{align*}
g(\delta) & =\left(-\delta^{2}+\left(1-\delta\right)C_{1}\right)\\
 & +\sqrt{\left(\delta^{2}+\left(1-\delta\right)C_{1}\right)^{2}+4\delta^{3}C_{1}}
\end{align*}
where $C_{1}=\frac{\sigma_{x}^{2}}{\sigma_{0}^{2}}$. Taking the derivative
of $g(\delta)$ with respect to $\delta$ and equating it to zero
results in $2C_{1}^{2}\delta^{2}+C_{1}^{3}\delta-2C_{1}^{3}=0$. We
thus have two saddle points
\[
\delta_{1}=\frac{-\sqrt{C_{1}^{2}+16C_{1}}-C_{1}}{4},\ \delta_{2}=\frac{\sqrt{C_{1}^{2}+16C_{1}}-C_{1}}{4}.
\]

Since $\delta$ must be non-negative, we consider only $\delta_{2}$.
In order to check whether $\delta_{2}$ is a minima, we rewrite $g(\delta)$
as
\begin{align*}
g(\delta) & =\sqrt{\left(\delta^{2}+\left(1-\delta\right)C_{1}\right)^{2}+4\delta^{3}C_{1}}-\delta^{2}+\left(1-\delta\right)C_{1}\\
 & \geq\sqrt{\left(\delta^{2}+\left(1-\delta\right)C_{1}\right)^{2}+4\delta^{3}C_{1}}-\left(\delta^{2}+\delta C_{1}\right)\\
 & =\sqrt{\left(\delta^{2}+\delta C_{1}\right)^{2}+2C_{1}\delta^{2}-2C_{1}^{2}\delta+C_{1}^{2}}-\left(\delta^{2}+\delta C_{1}\right).
\end{align*}
When $\delta\rightarrow\infty$, we have $2C_{1}\delta^{2}-2C_{1}^{2}\delta+C_{1}^{2}\rightarrow\infty$
which implies $g(\delta)\rightarrow\infty$, thus $\delta_{2}$ must
be a local minima. The optimal number of measurements is then
\begin{align}
\delta^{\dagger} & =\delta_{2}=\frac{\sqrt{C_{1}^{2}+16C_{1}}-C_{1}}{4}\label{eq:optimal delta Gaussian}\\
 & <\frac{\sqrt{C_{1}^{2}+16C_{1}+64}-C_{1}}{4}=\frac{\left(C_{1}+8\right)-C_{1}}{4}=2.\nonumber 
\end{align}
In addition, when
\begin{equation}
\sigma_{x}^{2}<2\sigma_{0}^{2},\label{eq:Gaussian_specific_condition}
\end{equation}
we have $C_{1}<2$. Since (\ref{eq:optimal delta Gaussian}) is a
monotonically increasing function, the condition $C_{1}<2$ results
in $\delta^{\dagger}<1$.
\end{IEEEproof}
The relationship between $\delta^{\dagger}$ and $C_{1}$ is simulated
in Fig. \ref{fig:Gaussian-Case:-Optimal}. The red solid line (Curve
1) is calculated via (\ref{eq:optimal delta Gaussian}) and the blue
dashed line (Curve 2) is achieved via ${\rm argmin_{\delta}}{\rm Err}$
where ${\rm Err}$ is the function of (\ref{eq:RandomMatrixTheory_Gaussian}). 
\begin{prop}[Least-favorable Distribution]
\label{claim:LF distribution}For a linear measurement system (\ref{eq:linear system 1})
with our proposed Gaussian noise model (\ref{eq:noise model}), assume
the signal elements are i.i.d. drawn from the least-favorable distribution
as in Section \ref{sec:Analysis-in-Real} for the real case and Section
\ref{sec:Analysis-in-Complex} for the complex case. Then $\delta^{\dagger}<2$.
When $M(\epsilon,\alpha^{\dagger})<0.5$ for the real case and $M_{C}(\epsilon,\alpha^{\dagger})<0.5$
for the complex case, we have $\delta^{\dagger}<1$.
\end{prop}
\begin{IEEEproof}
Based on Theorem \ref{thm:Theorem1}, we have $\delta^{\dagger}=2M(\epsilon,\alpha^{\dagger})$.
In order to bound $\delta^{\dagger}$, we need to consider a bound
on $M\left(\epsilon,\alpha\right)$. 

We first treat the real case, in which we rewrite $M\left(\epsilon,\alpha\right)$
as
\[
M\left(\epsilon,\alpha\right)=\epsilon T_{1}+T_{2},
\]
where
\begin{align*}
T_{1} & =\left(1+\alpha^{2}\right)+2\alpha\phi\left(\alpha\right)-2\left(1+\alpha^{2}\right)\Phi\left(-\alpha\right),\\
T_{2} & =2\left(1+\alpha^{2}\right)\Phi\left(-\alpha\right)-2\alpha\phi\left(\alpha\right).
\end{align*}
For any $\alpha\geq0$, we have $\Phi\left(-\alpha\right)\leq\frac{1}{2}$.
Thus,
\begin{align*}
T_{1} & =\left(1+\alpha^{2}\right)+2\alpha\phi\left(\alpha\right)-2\left(1+\alpha^{2}\right)\Phi\left(-\alpha\right)\\
 & \geq\left(1+\alpha^{2}\right)-2\left(1+\alpha^{2}\right)\Phi\left(-\alpha\right)\\
 & =\left(1+\alpha^{2}\right)\left(1-2\Phi\left(-\alpha\right)\right)\geq0
\end{align*}
which means that for any fixed $\alpha\geq0$, $M\left(\epsilon,\alpha\right)$
is a monotonically increasing function. Thus, for any $0<\epsilon_{1}<\epsilon_{2}\leq1$,
we have $M(\epsilon_{1},\alpha)<M(\epsilon_{2},\alpha)\leq M\left(1,\alpha\right)$,
where,
\[
M\left(1,\alpha\right)=1+\alpha^{2}.
\]

Let $\alpha_{1}^{\dagger}$ be the optimal value that minimises $M\left(1,\alpha\right)$.
Then $M(1,\alpha_{1}^{\dagger})=1$ and for any fixed $\alpha_{1}^{\dagger}$,
$M(\epsilon,\alpha_{1}^{\dagger})$ is a monotonically increasing
function as mentioned above, which means $M(\epsilon_{1},\alpha_{1}^{\dagger})<M(1,\alpha_{1}^{\dagger})$.
Let $\alpha_{\epsilon_{1}}^{\dagger}$ be the optimal value that minimises
$M(\epsilon_{1},\alpha)$. Then $M(\epsilon_{1},\alpha_{\epsilon_{1}}^{\dagger})\leq M(\epsilon_{1},\alpha_{1}^{\dagger})<1$
which means $M(\epsilon,\alpha^{\dagger})$ is upper bounded by $1$
and $\delta^{\dagger}=2M\left(\epsilon,\alpha^{\dagger}\right)$ is
upper bounded by $2$. Furthermore, for $M(\epsilon,\alpha^{\dagger})<0.5$,
we have $\delta^{\dagger}<1$ which only depends on $\epsilon$.

For the complex case, the analysis is similar. We rewrite (\ref{eq:Mc-fuction})
as $M_{C}\left(\epsilon,\alpha\right)=\epsilon T_{1}+T_{2}$ where
\begin{align*}
T_{1} & =1+\alpha^{2}-\sqrt{2\pi}\phi\left(\sqrt{2}\alpha\right)+2\alpha\sqrt{\pi}\Phi\left(-\sqrt{2}\alpha\right),\\
T_{2} & =\sqrt{2\pi}\phi\left(\sqrt{2}\alpha\right)-2\alpha\sqrt{\pi}\Phi\left(-\sqrt{2}\alpha\right).
\end{align*}
For any given $\alpha\geq0$, we have
\begin{align*}
T_{1} & \geq1+\alpha^{2}-\sqrt{2\pi}\phi\left(\sqrt{2}\alpha\right)\\
 & =1+\alpha^{2}-{\rm exp}\left(-\alpha^{2}\right)\geq0.
\end{align*}
By following the analysis in the real case, the same bound is achieved.
\end{IEEEproof}
To actually determine $\delta^{\dagger}$, we rely on numerical evaluations.
For any given $\epsilon$, there are no closed form formulas to compute
$\alpha^{\dagger}$ in (\ref{eq:optimal_alpha}) and $M\left(\epsilon,\alpha^{\dagger}\right)$
in (\ref{eq:M-function}): they have to be obtained numerically. As
a consequence, for any given value of $M\left(\epsilon,\alpha^{\dagger}\right)$,
the corresponding $\epsilon$ has to be found numerically. Simulations
in Fig. \ref{fig:M vs epsilon} show that for both real and complex
cases, $M\left(\epsilon,\alpha^{\dagger}\right)$ is upper bounded
by 1, and $\epsilon$ should be smaller (approximately) than $0.1928$
for the real case and $0.2289$ for the complex case to achieve $\delta^{\dagger}<1$.
\begin{prop}[Bernoulli-Gaussian Distribution]
\label{claim:BG distribution}For a linear measurement system (\ref{eq:linear system 1})
with our proposed Gaussian noise model (\ref{eq:noise model}), assume
the signal elements are i.i.d. drawn from the Bernoulli-Gaussian distribution
as in Section \ref{sec:Analysis-in-Real} for the real case and Section
\ref{sec:Analysis-in-Complex} for the complex case. Then $\delta^{\dagger}<2$
and when $\sigma_{x}^{2}<\frac{1}{\epsilon(1-R^{\infty}I\left(R^{\infty},\epsilon\right))}\sigma_{0}^{2}$
for the real case and $\sigma_{x}^{2}<\frac{1}{\epsilon(1-R^{\infty}I_{C}\left(R^{\infty},\epsilon\right))}\sigma_{0}^{2}$
for the complex case, we have $\delta^{\dagger}<1$.
\end{prop}
\begin{IEEEproof}
Based on Section \ref{subsec:Bernoulli-Gaussian-Distribution} we
known that
\begin{equation}
\delta^{\dagger}=\frac{\left(\sigma_{e}^{\infty}\right)^{2}}{2\sigma_{0}^{2}}\label{eq:OptimalDeltaBG}
\end{equation}
with the constraint $\left(\sigma_{e}^{\infty}\right)^{4}=4\sigma_{0}^{2}{\rm Err}_{\infty}\left(\epsilon\right)$.
For a fixed $\sigma_{0}^{2}$, we have $\delta^{\dagger}\propto\left(\sigma_{e}^{\infty}\right)^{2}\propto2\sigma_{0}\sqrt{{\rm Err}_{\infty}\left(\epsilon\right)}$,
thus we need to find the bound on ${\rm Err}_{\infty}\left(\epsilon\right)$
(here we use ${\rm Err}_{\infty}\left(\epsilon\right)$ instead of
${\rm Err}_{\infty}$ to highlight that ${\rm Err}_{\infty}\left(\epsilon\right)$
is a function of $\epsilon$).

Recall (\ref{eq:Err_BG}) which can be rewritten as
\begin{equation}
{\rm {\rm Err}_{\infty}}\left(\epsilon\right)=\sigma_{x}^{2}\epsilon-\sigma_{x}^{2}R^{\infty}I\left(R^{\infty},\epsilon\right)\epsilon\label{eq:Err_another_form}
\end{equation}
where $I\left(R^{\infty},\epsilon\right)$ is given by (\ref{eq:I1}).
For any given $R^{\infty}$ in (\ref{eq:Rt}) (which depends on $\sigma_{x}^{2}$
and $\sigma_{e}^{\infty}$ ), let $0<\epsilon_{1}<\epsilon_{2}\leq1$.
It is easy to verify that
\[
0<I\left(R^{\infty},\epsilon_{1}\right)<I\left(R^{\infty},\epsilon_{2}\right)\leq1.
\]
Define $f_{1}(\epsilon)\coloneqq\sigma_{x}^{2}\epsilon$ and $f_{2}\left(\epsilon\right)\coloneqq\sigma_{x}^{2}R^{\infty}I\left(R^{\infty},\epsilon\right)\epsilon$.
Then
\[
\frac{f_{1}(\epsilon)}{f_{2}\left(\epsilon\right)}=\frac{1}{R^{\infty}I\left(R^{\infty},\epsilon\right)}\geq1
\]
which means that ${\rm {\rm Err}_{\infty}}\left(\epsilon\right)$
is a monotonically increasing function of $\epsilon$ for any given
$\sigma_{x}^{2},$ $\sigma_{e}^{\infty}$ and $\sigma_{0}^{2}$. Thus
for $0<\epsilon_{1}<\epsilon_{2}\leq1$, we have
\[
0<{\rm Err}_{\infty}(\epsilon_{1})<{\rm Err}_{\infty}(\epsilon_{2})\leq{\rm Err}_{\infty}(1)
\]
and for $\epsilon=1$, the Bernoulli-Gaussian distribution degenerates
to the Gaussian signal. Thus ${\rm Err}_{\infty}\left(\epsilon\right)$
in the Bernoulli-Gaussian case is upper bounded by the Gaussian case,
in other words, $\delta^{\dagger}$ is upper bounded by $2$.

Consider the following condition
\begin{equation}
\sigma_{x}^{2}<\frac{1}{\epsilon(1-R^{\infty}I\left(R^{\infty},\epsilon\right))}\sigma_{0}^{2}.\label{eq:BG_specific_condition}
\end{equation}
Based on (\ref{eq:Err_another_form}) we have
\begin{align*}
{\rm {\rm Err}_{\infty}}\left(\epsilon\right) & <\sigma_{0}^{2}.
\end{align*}
Multiplying $4\sigma_{0}^{2}$ on both sides gives
\begin{align}
4\sigma_{0}^{2}{\rm Err}_{\infty}\left(\epsilon\right) & <4\sigma_{0}^{4}.\label{eq:cc}
\end{align}
For $\delta^{\dagger}$ we have the equality constraint $\left(\sigma_{e}^{\infty}\right)^{4}=4\sigma_{0}^{2}{\rm Err}_{\infty}\left(\epsilon\right)$.
Substituting $\left(\sigma_{e}^{\infty}\right)^{4}=4\sigma_{0}^{2}{\rm Err}_{\infty}\left(\epsilon\right)$
into (\ref{eq:cc}) results in
\begin{align*}
\frac{\left(\sigma_{e}^{\infty}\right)^{4}}{4\sigma_{0}^{4}} & <1.
\end{align*}
Taking the square root of both sides and only considering the real
value, we achieve the specific region
\[
\delta^{\dagger}=\frac{\left(\sigma_{e}^{\infty}\right)^{2}}{2\sigma_{0}^{2}}<1.
\]

The same analysis holds in the complex case. 
\end{IEEEproof}
Currently, there is no simple closed-form expression to describe the
relationship between $\sigma_{0}^{2}$, $\sigma_{x}^{2}$, $\epsilon$
and $\left(\sigma_{e}^{\infty}\right)^{2}$. Simulation of the specific
region $\delta^{\dagger}<1$ is provided in Fig. \ref{fig:Optimal_BG},
where $\delta^{\dagger}$ is calculated via (\ref{eq:OptimalDeltaBG}).
Here we set $\sigma_{x}^{2}=1$ and try different values of $\sigma_{e}^{\infty}$
such that $\left|\left(\sigma_{e}^{\infty}\right)^{4}-4\sigma_{0}^{2}{\rm Err}_{\infty}\right|<10^{-6}$,
after which we compute $\delta^{\dagger}=\frac{\left(\sigma_{e}^{\infty}\right)^{2}}{2\sigma_{0}^{2}}$
to find the optimal value. For the case $\epsilon=1$, we found $\delta^{\dagger}=1$
when $\sigma_{0}^{2}=0.5$ which coincides with the results from Fig.
\ref{fig:Our-Case}.

\subsection{Numerical Justification}

\begin{figure*}[tp]
\begin{centering}
\subfloat{\centering{}\includegraphics[scale=0.24]{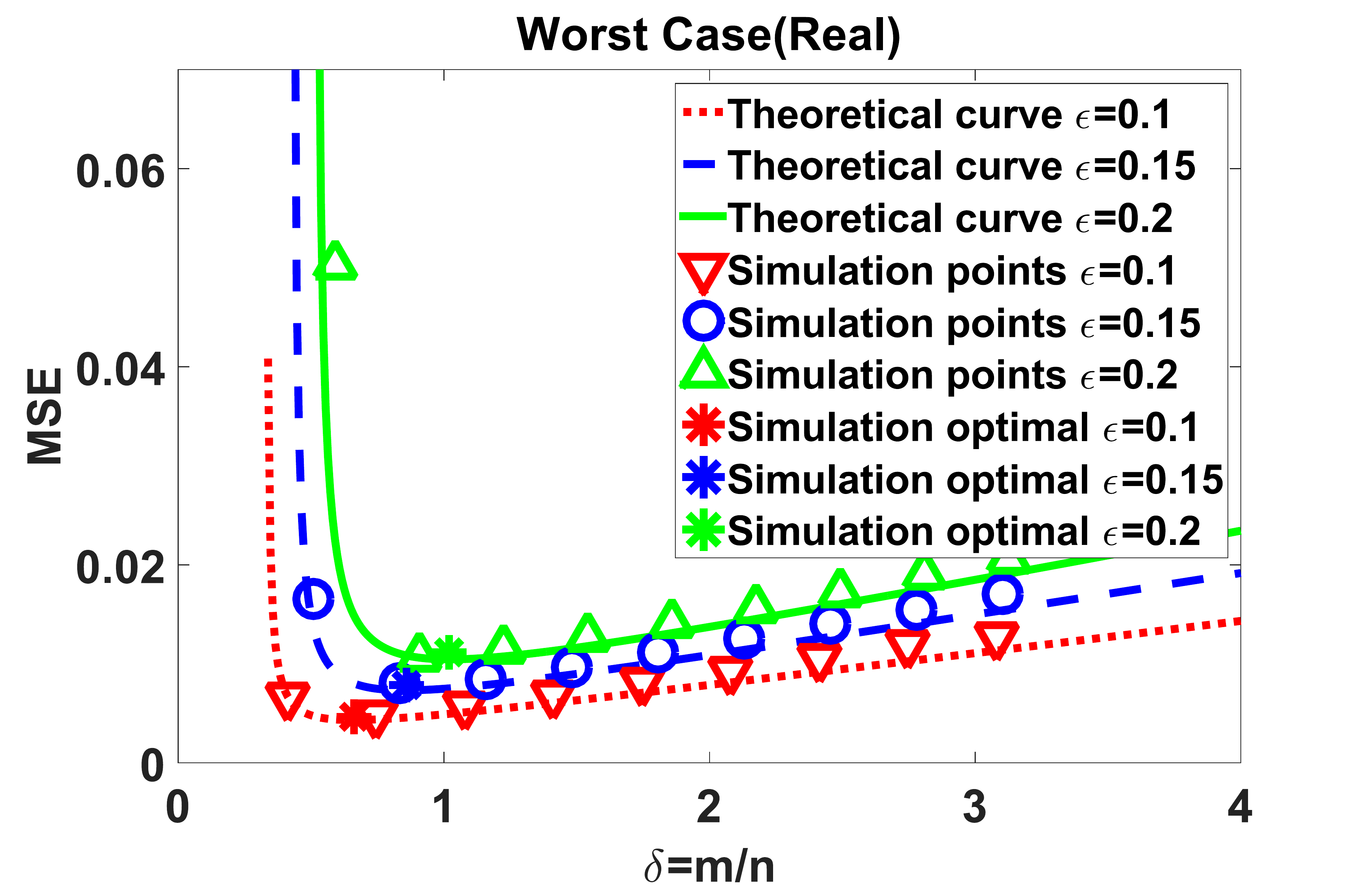}}\subfloat{\centering{}\includegraphics[scale=0.24]{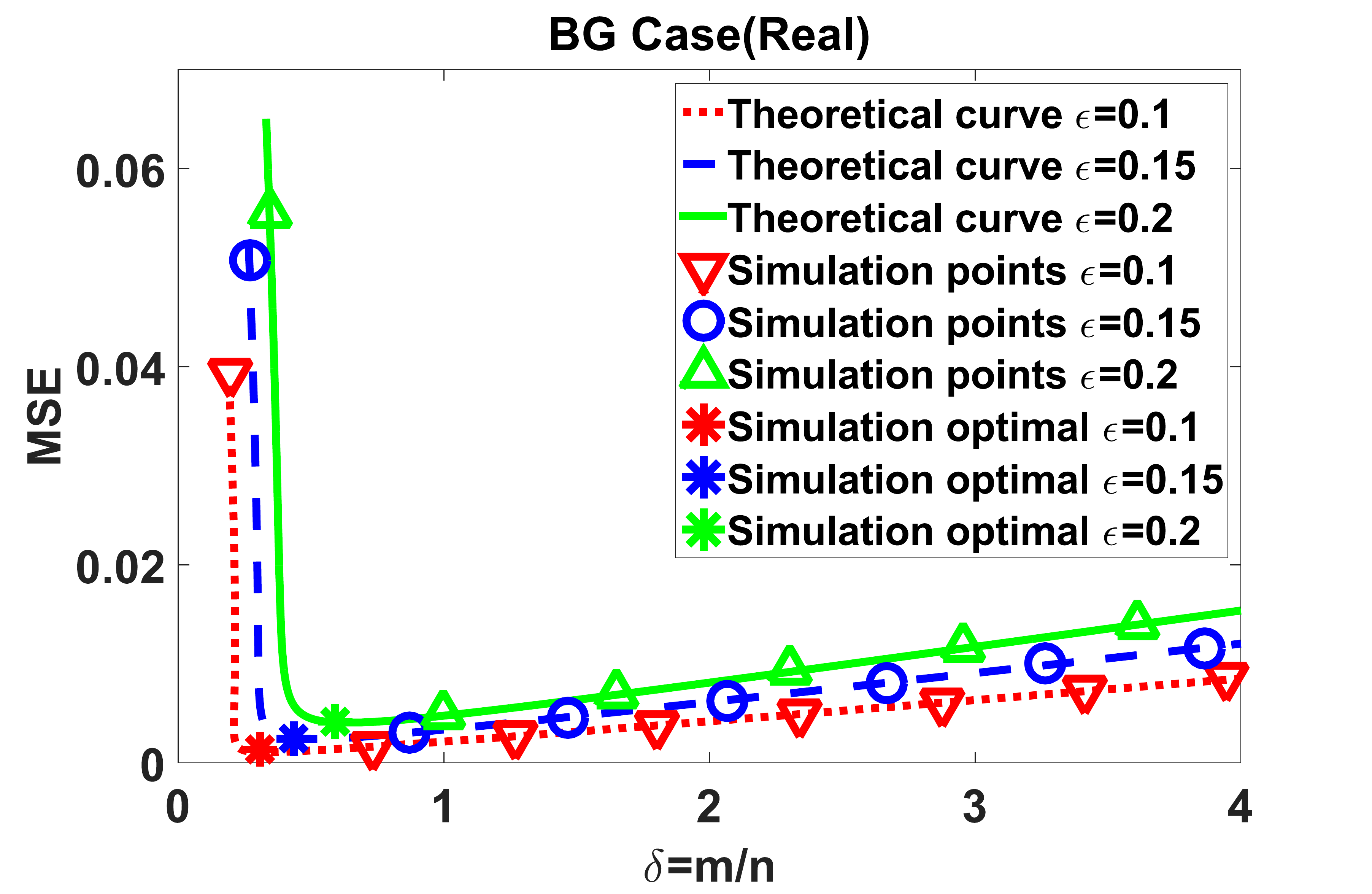}}
\par\end{centering}
\begin{centering}
\subfloat{\centering{}\includegraphics[scale=0.24]{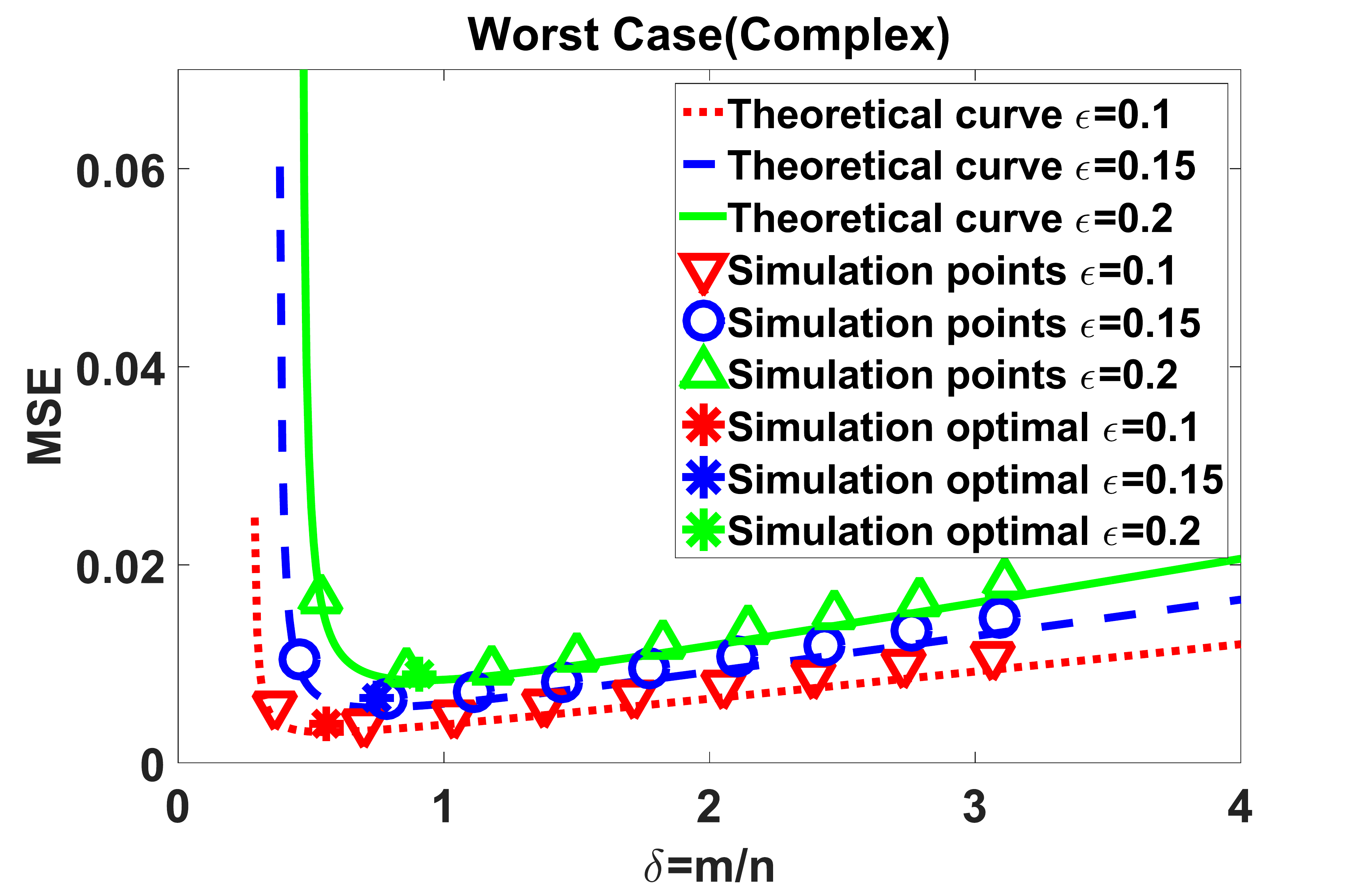}}\subfloat{\centering{}\includegraphics[scale=0.24]{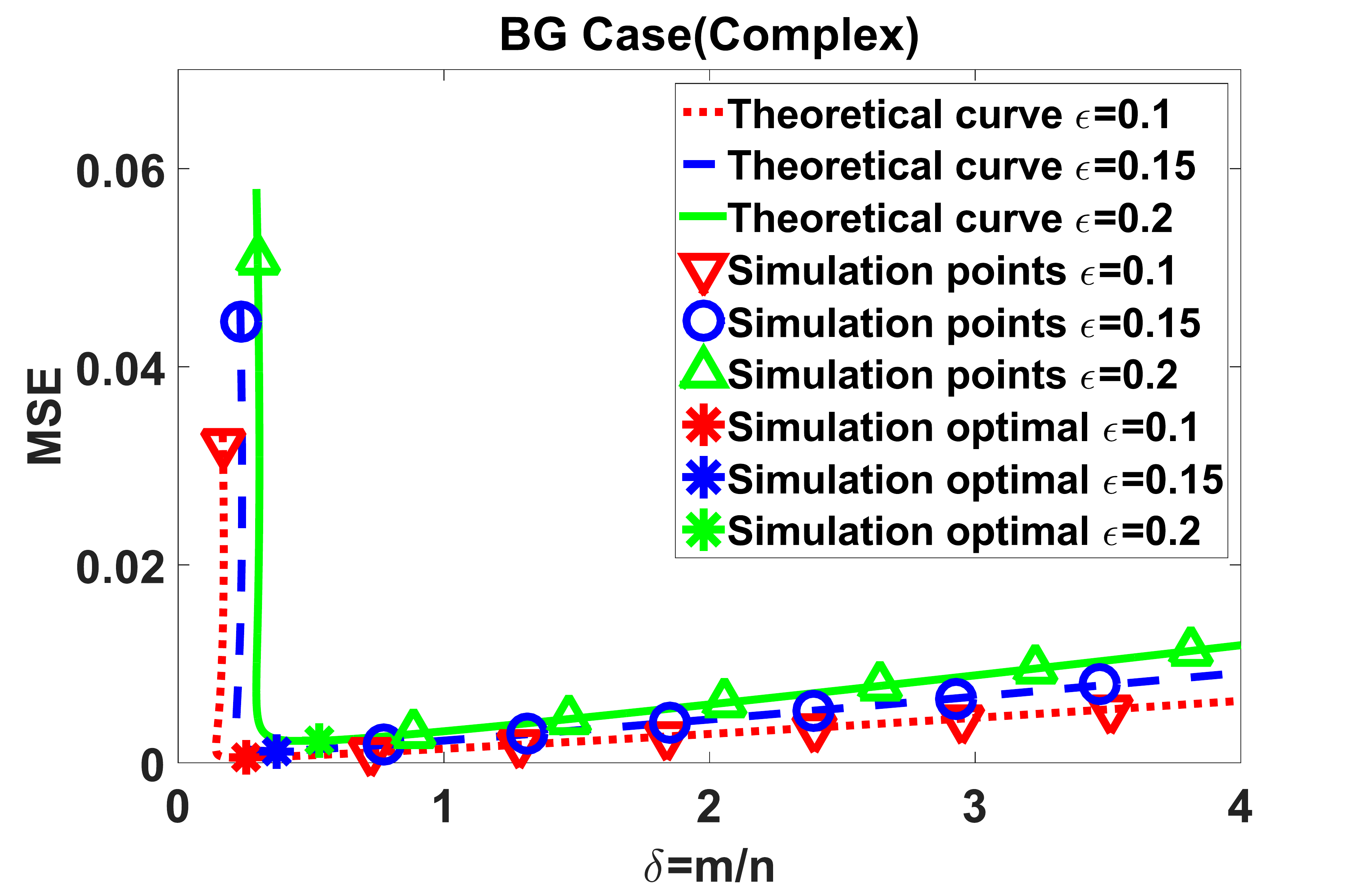}}
\par\end{centering}
\caption{\label{fig:MumericalSimulation}MSE (${\rm Err}_{\infty}$) vs $\delta$.}
\end{figure*}

Both the theorems provided in Section \ref{sec:Analysis-in-Real}
and the bounds in Section \ref{sec:BoundsANDSimulation} rely heavily
on the SE technique of the AMP algorithm. We compare the practical
curves of the MSE of AMP and the theoretical curves achieved by SE.

For the simulation, we set $n=1000$, $\sigma_{0}^{2}=0.01$ as constant
values. For the Bernoulli-Gaussian signal, we let $\sigma_{x}^{2}=1$,
and for the least-favorable signal, we use a Bernoulli-Gaussian distribution
but with large variance $\sigma_{x}^{2}=100$. Each simulation point
is the average of $100$ independent trials.

The simulation results provided in Fig. \ref{fig:MumericalSimulation}
show the relationship between the MSE and the measurement ratio $\delta$
for a given normalised sparsity level $\epsilon=\frac{S}{n}$. From
the figure, one can observe that when $\delta$ increases, the MSE
decreases dramatically until it reaches a minimum. After that, further
increase in $\delta$ increases the MSE. This phenomenon verifies
our presumption that there exists an optimal $\delta^{\dagger}$ (or
$m^{\dagger}$) for an energy fixed signal transmission system.

The overall performance of the Bernoulli-Gaussian distribution is
better than the least-favorable distribution. The numerical results
of the Bernoulli-Gaussian signal match the theoretical curves quite
well but for the least-favorable distribution, the numerical results
(MSE) are slightly larger than the theoretical curves. The main reason
is that for the theoretical analysis in this case, we assume that
the values of the non-zero coefficients are $\pm\infty$, but in simulations,
these values can only be set as certain large numbers which results
in a lower SNR compared with the one in the theoretical case. For
both signal distributions, the trends of $\delta^{\dagger}$ for different
$\epsilon$ values coincides with our theoretical analysis.

\section{Conclusions}

In this paper, we study the quadratically decreasing SNR setting by
assuming an energy limited system with a noise model whose variance
is proportional to the number of measurements. Analyses via random
matrix theory and state evolution both show there exists an optimal
choice of number of measurements to minimise the MSE of the estimated
signal. The obtained conclusion is quit different from the traditional
case which usually suggests the more measurements the better. Bounds
on the optimal value $\delta=\frac{m}{n}$ for three commonly used
signal distributions, Gaussian, Bernoulli-Gaussian and least-favorable,
show that in both the real and complex domains, the optimal value
$\delta$ is upper bounded by the value of 2. Specific situations
in which $\delta^{\dagger}<1$ for the three signal models have also
been analysed.

\appendix{}

\subsection{\label{sec:Proof(Real)}$\eta$ and ${\rm Err}_{t}$ for Bernoulli-Gaussian
Prior (Real)}

In this section, we provide the derivations of (\ref{eq:eta_BG_real})
and (\ref{eq:Err_BG}) in the real domain.

Assume the following scalar equation
\begin{equation}
\beta=x+w_{e}\label{eq:simple equation}
\end{equation}
where $x$ has density function (\ref{eq:BG-distribution}) and $w_{e}$
is the noise with density $\mathcal{N}\left(0,\sigma_{e}^{2}\right)$.
The joint probability of $x$ and $\beta$ is
\begin{align*}
p\left(x,\beta\right)= & p_{G}\left(\beta-x;\,0,\,\sigma_{e}^{2}\right)\left(1-\epsilon\right)\Delta_{x=0}\\
 & +\epsilon p_{G}\left(\beta-x;\,0,\,\sigma_{e}^{2}\right)p_{G}\left(x;\,0,\,\sigma_{x}^{2}\right)
\end{align*}
and
\[
p\left(\beta\right)=\left(1-\epsilon\right)p_{G}\left(\beta;\,0,\,\sigma_{e}^{2}\right)+\epsilon p_{G}\left(\beta;\,0,\,\sigma_{x}^{2}+\sigma_{e}^{2}\right).
\]
 Let the estimation be $\hat{x}=\eta\left(\beta\right)={\rm E}_{x|\beta}\left[x|\beta\right]$.
Based on Bayes' theorem, we have
\begin{align}
\hat{x} & =\int x\frac{p\left(x,\beta\right)}{p\left(\beta\right)}dx\nonumber \\
 & =\frac{\int xp_{G}\left(\beta-x;\,x,\,\sigma_{e}^{2}\right)p_{G}\left(x;0,\,\sigma_{x}^{2}\right)dx}{\frac{\left(1-\epsilon\right)}{\epsilon}p_{G}\left(\beta;\,0,\,\sigma_{e}^{2}\right)+p_{G}\left(\beta;\,0,\,\sigma_{x}^{2}+\sigma_{e}^{2}\right)}.\label{eq:simple mean}
\end{align}
We next rely on the following lemma.
\begin{lem}
\label{lem:GaussianConditionalLemma}Let $\bm{y}=\bm{A}\bm{x}+\bm{w}$
where $\bm{A}\in\mathbb{R}^{m\times n}$ is a deterministic matrix,
$\bm{x}\sim\mathcal{N}\left(\bm{0},\bm{\Sigma}_{x}\right)$ and $\bm{w}\sim\mathcal{N}\left(\bm{0},\bm{\Sigma}_{w}\right)$
are Gaussian random vectors. Assume all the matrix inverses exist.
Then
\begin{align*}
\bm{\mu}_{\bm{x}|\bm{y}} & =\bm{\Sigma}_{x}\bm{A}^{T}\left(\bm{A}\bm{\Sigma}_{x}\bm{A}^{T}+\bm{\Sigma}_{w}\right)^{-1}\bm{y}\\
 & =\left(\bm{\Sigma}_{x}^{-1}+\bm{A}^{T}\bm{\Sigma}_{w}^{-1}\bm{A}\right)^{-1}\bm{A}^{T}\bm{\Sigma}_{w}^{-1}\bm{y},\\
\bm{\Sigma}_{\bm{x}|\bm{y}} & =\left(\bm{\Sigma}_{x}^{-1}+\bm{A}^{T}\bm{\Sigma}_{w}^{-1}\bm{A}\right)^{-1}\\
 & =\bm{\Sigma}_{x}-\bm{\Sigma}_{x}\bm{A}^{T}\left(\bm{A}\bm{\Sigma}_{x}\bm{A}^{T}+\bm{\Sigma}_{w}\right)^{-1}\bm{\Sigma}_{x}
\end{align*}
where $\bm{\mu}_{\bm{x}|\bm{y}}$ is the conditional mean and $\bm{\Sigma}_{\bm{x}|\bm{y}}$
is the conditional covariance matrix.
\end{lem}
Consider the mean value $\bm{\mu}_{\bm{x}|\bm{y}}$ in Lemma \ref{lem:GaussianConditionalLemma},
which is used to compute ${\rm E}_{\bm{x}|\bm{y}}\left[\bm{x}|\bm{y}\right]$.
Now apply this result to the scalar function (\ref{eq:simple equation}).
By setting $\epsilon=1$, we have the following fact
\begin{align}
\frac{\int xp_{G}\left(\beta-x;\,x,\,\sigma_{e}^{2}\right)p_{G}\left(x;0,\,\sigma_{x}^{2}\right)dx}{p_{G}\left(\beta;\,0,\,\sigma_{x}^{2}+\sigma_{e}^{2}\right)} & =R\beta,\label{eq:fact}
\end{align}
where $R\coloneqq\frac{\sigma_{x}^{2}}{\sigma_{x}^{2}+\sigma_{e}^{2}}$.
Substituting (\ref{eq:fact}) into (\ref{eq:simple mean}) provides
\begin{align*}
\hat{x} & =\frac{p_{G}\left(\beta;\,0,\,\sigma_{x}^{2}+\sigma_{e}^{2}\right)\epsilon R\beta}{\left(1-\epsilon\right)p_{G}\left(\beta;\,0,\,\sigma_{e}^{2}\right)+\epsilon p_{G}\left(\beta;\,0,\,\sigma_{x}^{2}+\sigma_{e}^{2}\right)}\\
 & =\frac{p_{G}\left(\beta;\,0,\,\sigma_{x}^{2}+\sigma_{e}^{2}\right)\epsilon R\beta}{p\left(\beta\right)}
\end{align*}
which has the same form as (\ref{eq:eta_BG_real}).

Now consider the MSE as ${\rm Err}={\rm E}\left[\left(x-\hat{x}\right)^{2}\right]$.
Thus
\begin{align*}
{\rm Err} & ={\rm E}\left[x^{2}-2{\rm E}_{x|\beta}\left[x|\beta\right]x+{\rm E}_{x|\beta}\left[x|\beta\right]^{2}\right]\\
 & ={\rm E}_{\beta}\left[{\rm E}_{x|\beta}\left[x^{2}-2{\rm E}_{x|\beta}\left[x|\beta\right]x+{\rm E}_{x|\beta}\left[x|\beta\right]^{2}\right]\right]\\
 & ={\rm E}_{\beta}\left[{\rm E}_{x|\beta}\left[x^{2}|\beta\right]\right]-{\rm E}_{\beta}\left[{\rm E}_{x|\beta}\left[x|\beta\right]^{2}\right].
\end{align*}
Note that ${\rm E}_{\beta}\left[{\rm E}_{x|\beta}\left[x^{2}|\beta\right]\right]={\rm E}\left[x^{2}\right]=\epsilon\sigma_{x}^{2}$
based on the law of total expectation. For ${\rm E}_{\beta}\left[{\rm E}_{x|\beta}\left[x|\beta\right]^{2}\right]$
we have
\begin{align*}
{\rm E}_{\beta}\left[{\rm E}_{x|\beta}\left[x|\beta\right]^{2}\right]= & \int\frac{p_{G}\left(\beta;\,0,\,\sigma_{x}^{2}+\sigma_{e}^{2}\right)^{2}\epsilon^{2}R^{2}\beta^{2}}{p\left(\beta\right)}d\beta\\
= & \epsilon\sigma_{x}^{2}R\int\frac{\frac{1}{\sigma_{x}^{2}+\sigma_{e}^{2}}p_{G}\left(\beta;\,0,\,\sigma_{x}^{2}+\sigma_{e}^{2}\right)\beta^{2}}{\frac{\left(1-\epsilon\right)}{\epsilon}\frac{p_{G}\left(\beta;\,0,\,\sigma_{e}^{2}\right)}{p_{G}\left(\beta;\,0,\,\sigma_{x}^{2}+\sigma_{e}^{2}\right)}+1}d\beta.
\end{align*}
Define $\beta\coloneqq\sqrt{\sigma_{x}^{2}+\sigma_{e}^{2}}\gamma$.
Then{\small{}
\begin{align*}
{\rm E}_{\beta}\left[{\rm E}_{x|\beta}\left[x|\beta\right]^{2}\right]= & \epsilon\sigma_{x}^{2}R\int\frac{\frac{1}{\sqrt{2\pi}}{\rm exp}\left(-\frac{\gamma^{2}}{2}\right)}{\frac{\left(1-\epsilon\right)}{\epsilon}\sqrt{\frac{1}{1-R}}{\rm exp}\left(-\frac{\gamma^{2}\sigma_{x}^{2}}{2\sigma_{e}^{2}}\right)+1}\gamma^{2}d\gamma\\
= & \epsilon\sigma_{x}^{2}RI\left(R,\epsilon\right)
\end{align*}
}and
\begin{align*}
{\rm Err} & =\epsilon\sigma_{x}^{2}-\epsilon\sigma_{x}^{2}RI\left(R,\epsilon\right)\\
 & =\left[\frac{R\epsilon}{1-R}\left(1-RI\left(R,\epsilon\right)\right)\right]\sigma_{e}^{2}
\end{align*}
which has the same form as (\ref{eq:Err_BG}).

\subsection{\label{subsec:Proof(Complex)}$\eta$ and ${\rm Err}_{t}$ for Bernoulli-Gaussian
Prior (Complex)}

In this section, we extend (\ref{eq:eta_BG_real}) and (\ref{eq:Err_BG})
to the complex case. We start from (\ref{eq:complex Gaussian}). Following
the same steps as in Appendix \ref{sec:Proof(Real)}, we have
\begin{align*}
\hat{x} & ={\rm E}\left[x|\beta\right]=\int xp\left(x|\beta\right)dx,\\
 & =\frac{p_{CG}\left(\beta;0,\sigma_{e}^{2}+\sigma_{x}^{2}\right)}{p_{C}\left(\beta\right)}\epsilon R\beta,
\end{align*}
where
\[
p_{C}\left(\beta\right)=\left(1-\epsilon\right)p_{CG}\left(\beta;0,\sigma_{e}^{2}\right)+\epsilon p_{CG}\left(\beta;0,\sigma_{e}^{2}+\sigma_{x}^{2}\right).
\]
The MSE calculation in the complex case, becomes
\begin{align*}
{\rm Err}_{C} & ={\rm E}x^{2}-{\rm E}_{\beta}\left[\left|{\rm E}_{x|\beta}\left[x|\beta\right]\right|^{2}\right]
\end{align*}
where we still have ${\rm E}x^{2}=\epsilon\sigma_{x}^{2}.$ The second
term gives
\begin{align}
 & {\rm E}_{\beta}\left[\left|{\rm E}_{x|\beta}\left[x|\beta\right]\right|^{2}\right]\nonumber \\
 & =\int\left|E_{x|\beta}\left[x|\beta\right]\right|^{2}p_{C}\left(\beta\right)d\beta\\
 & =\epsilon R^{2}\int_{\beta^{R}}\int_{\beta^{I}}\frac{p_{CG}\left(\beta;0,\sigma_{e}^{2}+\sigma_{x}^{2}\right)}{\frac{\left(1-\epsilon\right)}{\epsilon}\frac{p_{CG}\left(\beta;0,\sigma_{e}^{2}\right)}{p_{CG}\left(\beta;0,\sigma_{e}^{2}+\sigma_{x}^{2}\right)}+1}\left|\beta\right|^{2}d\beta^{R}d\beta^{I}.\label{eq:err2complex}
\end{align}
In addition, we have $p_{CG}\left(x;0,\sigma_{x}^{2}\right)=\frac{1}{\sigma_{x}^{2}}\frac{1}{\pi}{\rm exp}\left(-\frac{\left|x\right|^{2}}{\sigma_{x}^{2}}\right)$.

Let $x=\sigma_{x}y$. Then,
\begin{align*}
p_{CG}\left(x;0,\sigma_{x}^{2}\right) & =\frac{1}{\sigma_{x}^{2}}\frac{1}{\pi}{\rm exp}\left(-\frac{\sigma_{x}^{2}\left|y\right|^{2}}{\sigma_{x}^{2}}\right)\\
 & =\frac{1}{\sigma_{x}^{2}}\phi_{c}\left(y\right)=\frac{1}{\sigma_{x}^{2}}\phi_{c}\left(\frac{x}{\sigma_{x}}\right),
\end{align*}
which implies that we can rewrite (\ref{eq:err2complex}) as
\begin{align}
 & {\rm E}_{\beta}\left[\left|{\rm E}_{x|\beta}\left[x|\beta\right]\right|^{2}\right]\nonumber \\
 & =\epsilon R^{2}\int_{\beta^{R}}\int_{\beta^{I}}\frac{\frac{1}{\sigma_{e}^{2}+\sigma_{x}^{2}}\phi_{c}\left(\frac{\beta}{\sqrt{\sigma_{e}^{2}+\sigma_{x}^{2}}}\right)}{\frac{\left(1-\epsilon\right)}{\epsilon}\frac{1}{1-R}{\rm exp}\left(-R\frac{\left|\beta\right|^{2}}{\sigma_{e}^{2}}\right)+1}\left|\beta\right|^{2}d\beta^{R}d\beta^{I}.\label{eq:err2complex2}
\end{align}
Define $\beta\coloneqq\sqrt{\sigma_{x}^{2}+\sigma_{e}^{2}}\gamma$.
We then have $d\beta^{R}=\sqrt{\sigma_{e}^{2}+\sigma_{x}^{2}}d\gamma^{R}$
, $d\beta^{I}=\sqrt{\sigma_{e}^{2}+\sigma_{x}^{2}}d\gamma^{I}$, and
$\left|\beta\right|^{2}=\left(\sigma_{e}^{2}+\sigma_{x}^{2}\right)\left|\gamma\right|^{2}$.
Substituting into (\ref{eq:err2complex2}) leads to
\begin{align*}
 & {\rm Err}_{C}\\
 & =\epsilon\sigma_{x}^{2}\\
 & -\sigma_{x}^{2}\epsilon R\int_{\gamma^{R}}\int_{\gamma^{I}}\frac{\phi_{c}\left(\gamma\right)}{\frac{\left(1-\epsilon\right)}{\epsilon}\frac{1}{1-R}{\rm exp}\left(-\frac{R}{1-R}\left|\gamma\right|^{2}\right)+1}\left|\gamma\right|^{2}d\gamma^{R}d\gamma^{I},\\
 & =\epsilon\sigma_{x}^{2}\left(1-RI_{C}\left(R,\epsilon\right)\right)=\left[\frac{R\epsilon}{1-R}\left(1-RI_{C}\left(R,\epsilon\right)\right)\right]\sigma_{e}^{2}
\end{align*}
which has the same form as (\ref{eq:Err_BG_c}).

\bibliographystyle{IEEEtran}
\bibliography{FinalVersion}

\end{document}